\documentclass[11pt]{article}
\usepackage[T1]{fontenc}
\usepackage[utf8]{inputenc}
\usepackage{lmodern}
\usepackage{microtype}
\usepackage[margin=1in]{geometry}
\usepackage{amsmath,amssymb}
\usepackage{graphicx}
\usepackage{longtable}
\usepackage{placeins}
\usepackage[numbers,sort&compress]{natbib}
\usepackage{authblk}
\usepackage[font=small,labelfont=bf]{caption}
\usepackage{booktabs}
\usepackage{multirow}
\usepackage{array}
\usepackage{pifont}
\newcommand{\cmark}{\ding{51}}
\newcommand{\xmark}{\ding{55}}
\newcommand{\pmark}{$\triangle$}
\usepackage[colorlinks=true,allcolors=blue]{hyperref}

\title{Privacy-preserving Machine Learning in Internet of Vehicle Applications: Fundamentals, Recent Advances, and Future Directions}

\author[1]{Nazmul Islam\thanks{Corresponding author. E-mail: \href{mailto:nazmul.islam@queensu.ca}{nazmul.islam@queensu.ca}}}
\author[1]{Mohammad Zulkernine}
\affil[1]{School of Computing, Queen's University, Kingston, ON K7L 3N6, Canada\authorcr
\upshape\texttt{\{nazmul.islam, mz\}@queensu.ca}}
\date{}

\begin{document}

\maketitle

\begin{abstract}
Machine learning (ML) in Internet of Vehicles (IoV) applications enhanced intelligent transportation, autonomous driving capabilities, and various connected services within a large, heterogeneous network. However, the increased connectivity and massive data exchange for ML applications introduce significant privacy challenges. Privacy-preserving machine learning (PPML) offers potential solutions to address these challenges by preserving privacy at various stages of the ML pipeline. Despite the rapid development of ML-based IoV applications and the growing data privacy concerns, there are limited comprehensive studies on the adoption of PPML within this domain. Therefore, this study provides a comprehensive review of the fundamentals, recent advancements, and the challenges of integrating PPML into IoV applications. We first review existing surveys of various PPML techniques and their integration into IoV across different scopes. We then categorize IoV applications into three key domains and analyze the privacy challenges in leveraging ML in these application domains. Building on these fundamentals, we review recent advancements in integrating various PPML techniques within IoV applications, discussing their frameworks, key features, and performance. Finally, we identify current challenges and propose future research directions to enhance privacy and reliability in IoV applications.

\end{abstract}

\section{Introduction}\label{sec:intro}

The internet of vehicles (IoV) integrates advanced communication technologies and intelligent systems to establish a connected ecosystem of vehicles, infrastructure, and services. Leveraging real-time data, IoV systems can predict traffic patterns, optimize routes, and dynamically adjust traffic signals for improved transportation efficiency. For autonomous driving, these systems integrate advanced sensing and computer vision, enabling vehicles to perceive their surroundings, make informed decisions, and navigate safely without human intervention. Furthermore, IoV systems integrate with urban ecosystem infrastructure to enable smart services such as intelligent parking, dynamic electric vehicle (EV) charging, and personalized infotainment.

IoV inherently generates vast and diverse amounts of data from vehicles, infrastructure, pedestrians, and other sources in the ecosystem. Given the data-intensive nature of machine learning (ML) algorithms and their reliance on large-scale datasets for training, these data form an ideal foundation for ML applications \cite{ref4}. This led to the rapid adoption of ML in IoV applications over the past decade. The integration of ML into IoV has significantly improved existing applications in the ecosystem, while also enabling new capabilities such as autonomous driving and smart EV charging. However, IoV data includes sensitive and private information such as location details, driving patterns, and personally identifiable information, raising significant privacy concerns regarding its collection, use, and potential misuse or breaches \cite{ref4,ref6}.

In ML-based IoV applications, various attacks can compromise both the training and inference stages of the ML pipeline, as well as the data collection and storage processes \cite{ref7,ref8}. During the training phase, attacks such as data poisoning, property inference, or Byzantine attacks, and in the inference phase, membership inference, model inversion, or model extraction attacks can undermine the privacy and integrity of both the model and its underlying data \cite{ref8,ref9}. Moreover, attacks targeting communication channels, such as man-in-the-middle, can intercept or alter sensitive data, compromising data security and integrity. The involvement of third-party services in data collection, model training, or deployment introduces additional vulnerabilities, as mishandling of data or breaches at third-party entities can threaten the entire IoV ecosystem and compromise vehicle security and user privacy on a broader scale \cite{ref10}.

To circumvent these privacy risks, advanced privacy-preserving ML (PPML) techniques such as federated learning (FL), homomorphic encryption (HE), secure multi-party computations (SMPC), and differential privacy (DP) are being thoroughly studied to protect user data and model integrity in the ML pipeline \cite{ref8,ref9,ref10}. Furthermore, blockchain-based PPML (BC-PPML) enhances security and trust in distributed IoV systems while preserving privacy.

\subsection{Existing Surveys}

For completeness of the study, we have considered two main areas of related research as summarized in Table~\ref{tbl:related-work}. First, studies that cover various PPML techniques and second, studies that review privacy challenges in specific IoV application domains or PPML techniques within a domain.

Several studies have presented the theoretical and fundamental aspects of PPML techniques. For instance, studies \cite{ref7,ref8}, \cite{ref9,ref10} reviewed privacy issues in deep learning (DL), including attacks, privacy-preserving techniques and their challenges. In the context of distributed systems, \cite{ref11,ref12} surveyed distributed DL alongside privacy-preservation techniques and their optimization. Furthermore, \cite{ref13} provided a comprehensive taxonomy and structured overview of privacy-preserving DL (PPDL). Besides broader survey studies, there are studies that focus on specific PPML techniques such as DP techniques for cyber-physical systems \cite{ref14}, privacy preservation in BC-based IoT systems \cite{ref15}, DP in deep-FL \cite{ref16}, and HE \cite{ref17} and SMPC \cite{ref18} for secure PPML computation. Furthermore, FL is a key PPML technique and widely studied, as reviewed in \cite{ref19,ref20,ref21} on its taxonomy, advancements, and security challenges. While these studies provided key insights into the theoretical foundations and applications of PPML, they do not cover domain specific challenges or requirements like the IoV.

In the context of IoV, several survey papers provided broad overview of privacy and security challenges such as security in intelligent transportation systems (ITS) \cite{ref22,ref23}, ML for vehicular network security \cite{ref6}, PPML for electric vehicles (EVs) \cite{ref24}, and security in 6G-enabled IoV \cite{ref25}. A few studies focused on specific PPML techniques within IoV applications, such as HE in vehicular ad hoc networks (VANETs) \cite{ref26}, local-DP for secure IoV \cite{ref27}, BC-PPML in IoV \cite{ref28}, BC-enabled FL in IoV \cite{ref29}, BC for IoV (discussed some aspects of BC-PPML) \cite{ref30}, and FL in connected autonomous vehicles \cite{ref31} and ITS applications \cite{ref33,ref32}.

\begin{table}[htbp]
\centering
\scriptsize\normalfont
\setlength{\tabcolsep}{2pt}
\setlength{\aboverulesep}{0pt}
\setlength{\belowrulesep}{0pt}
\caption{Summary and comparison of existing literature reviews.}\label{tbl:related-work}
\begin{tabular}{@{}>{\centering\arraybackslash}p{0.08\linewidth}p{0.48\linewidth}*{5}{>{\centering\arraybackslash}m{0.035\linewidth}}*{3}{>{\centering\arraybackslash}m{0.06\linewidth}}@{}}
\toprule
\multirow{2}{*}{\textbf{Paper*}} & \multirow{2}{0.58\linewidth}{\textbf{Focus of the Study}} & \multicolumn{5}{@{}c@{}}{\textbf{PPML Techniques**}} & \multicolumn{3}{@{}c@{}}{\textbf{IoV Applications**}} \\
\cmidrule(l){3-7}\cmidrule(lr){8-10} & & \textbf{T1} & \textbf{T2} & \textbf{T3} & \textbf{T4} & \textbf{T5} & \textbf{D1} & \textbf{D2} & \textbf{D3} \\
\midrule
\cite{ref8,ref9,ref10,ref11,ref13} & Privacy in DL, including a review of PPML techniques, taxonomy, methods, and challenges and future direction & $\bullet$ & - & $\bullet$ & $\bullet$ & $\bullet$ & - & - & - \\
\cite{ref7} & Privacy attacks in machine learning & $\bullet$ & - & - & - & $\bullet$ & - & - & - \\
\cite{ref12} & Privacy-preserving distributed optimization and learning & - & - & $\bullet$ & $\bullet$ & $\bullet$ & - & - & - \\
\cite{ref14} & Differential privacy techniques in cyber physical system & - & - & - & - & $\bullet$ & - & - & $\bullet$ \\
\cite{ref15} & Privacy-preserving blockchain-IoT integration & - & $\bullet$ & - & - & - & - & - & $\circ$ \\
\cite{ref16} & Differential privacy in deep and FL & $\bullet$ & - & - & - & $\bullet$ & - & - & - \\
\cite{ref17} & Privacy-preserving machine learning with fully HE & - & - & $\bullet$ & - & - & - & - & - \\
\cite{ref18} & Privacy-preserving deep learning with SMPC & - & - & - & $\bullet$ & - & - & - & - \\
\cite{ref19} & Taxonomy and review of privacy-preserving FL & $\bullet$ & - & - & - & - & - & - & - \\
\cite{ref20} & Fundamentals, state of the art, trends, and challenges in DFL & $\bullet$ & - & - & - & - & - & - & $\circ$ \\
\cite{ref21} & Privacy-preserving and secure robust FL techniques & $\bullet$ & - & $\bullet$ & - & $\bullet$ & - & - & - \\
\midrule
\cite{ref22} & Classification of security and privacy vulnerabilities in ITS & - & - & - & - & - & $\bullet$ & - & - \\
\cite{ref23} & ITS security challenges and potential solutions & - & - & - & - & - & $\bullet$ & - & - \\
\cite{ref6} & Machine learning for security in vehicular networks & $\circ$ & - & - & - & - & $\bullet$ & - & - \\
\cite{ref24} & Privacy-preservation techniques in electric vehicles & $\bullet$ & $\circ$ & $\circ$ & - & $\bullet$ & - & $\bullet$ & - \\
\cite{ref25} & Security and privacy framework for 6G vehicular networks & - & - & - & - & - & - & - & $\bullet$ \\
\cite{ref26} & Secure computation methods based on HE in VANETs & - & - & $\bullet$ & - & - & - & - & $\bullet$ \\
\cite{ref27} & Local DP techniques for securing IoVs & - & - & - & - & $\bullet$ & $\bullet$ & - & $\bullet$ \\
\cite{ref28} & Privacy-preserving solutions using blockchain for VANETs & - & $\bullet$ & - & - & - & $\bullet$ & $\circ$ & $\bullet$ \\
\cite{ref29} & Analysis of blockchain-enabled federated learning & $\bullet$ & $\bullet$ & - & - & - & $\bullet$ & $\circ$ & $\circ$ \\
\cite{ref30} & Blockchain intelligence for IoV, with challenges and solutions & - & $\bullet$ & - & - & - & $\bullet$ & $\bullet$ & $\bullet$ \\
\cite{ref31} & Review of FL approaches in vehicles & $\bullet$ & $\circ$ & - & - & - & - & $\bullet$ & $\circ$ \\
\cite{ref32} & Recent applications and open problems in FL for ITS & $\bullet$ & $\circ$ & - & - & $\circ$ & $\bullet$ & - & $\circ$ \\
\cite{ref33} & Analysis of FL applications in ITS & $\bullet$ & $\circ$ & - & - & $\circ$ & $\bullet$ & - & $\circ$ \\
\midrule
\multicolumn{2}{@{}p{0.56\linewidth}}{\textbf{This Study.} Overview of IoV ecosystem, its privacy concerns, and the integration of PPML techniques in IoV applications.} & $\bullet$ & $\bullet$ & $\bullet$ & $\bullet$ & $\bullet$ & $\bullet$ & $\bullet$ & $\bullet$ \\
\bottomrule
\multicolumn{10}{@{}p{\linewidth}@{}}{\scriptsize *The top group consists of surveys on the fundamentals of PPML, while the bottom group is on the integration of PPML in IoV applications. **The notations indicate topics fully covered ($\bullet$), partially covered ($\circ$), and not covered (-) in the study. The PPML techniques are T1: FL, T2: BC-PPML, T3: HE, T4: SMPC and T5: DP. The application domains are D1: Intelligent transportation and traffic management, D2: Autonomous driving and safety-critical applications and D3: Communication infrastructure and smart services.} \\
\end{tabular}
\end{table}

\subsection{Scope and Contribution}

Existing survey literature generally covered either a broad review of PPML technologies, security and privacy in IoV, or specific PPML techniques within a narrow application domain. The current study provides a comprehensive survey of PPML techniques in vast IoV applications. To achieve this, we first provide an overview of the IoV ecosystem with proposed architecture and then categorize the applications into three major domains: (1) intelligent transportation and traffic management, (2) autonomous driving and safety-critical applications, and (3) communication infrastructure and smart services. We then analyze the various data types within each domain and their associated privacy challenges. Following this, we review various PPML techniques and compare their performance and applicability. Building on these foundations, we review the recent advancements in the adoption of key PPML techniques within the three application domains focusing on their implementation and performance evaluation in terms of privacy, utility and efficiency. Finally, we identify gaps in current research and propose future directions for the integration of PPML in IoV.

The scope of this survey is privacy-preserving techniques employed within ML processes in IoV applications. Notably, our survey excludes studies adopting privacy-enhancing techniques such as data perturbation, anonymization, pseudonymization, k-anonymity, and others, unless they are specifically integrated into PPML for data privacy in IoV applications. Additionally, techniques related to authentication, authorization, and access control, aimed at securing user identities or access to data, are out of scope.

\subsection{Survey Structure}

\begin{figure}[htbp]
  \centering
  \includegraphics[width=0.92\textwidth]{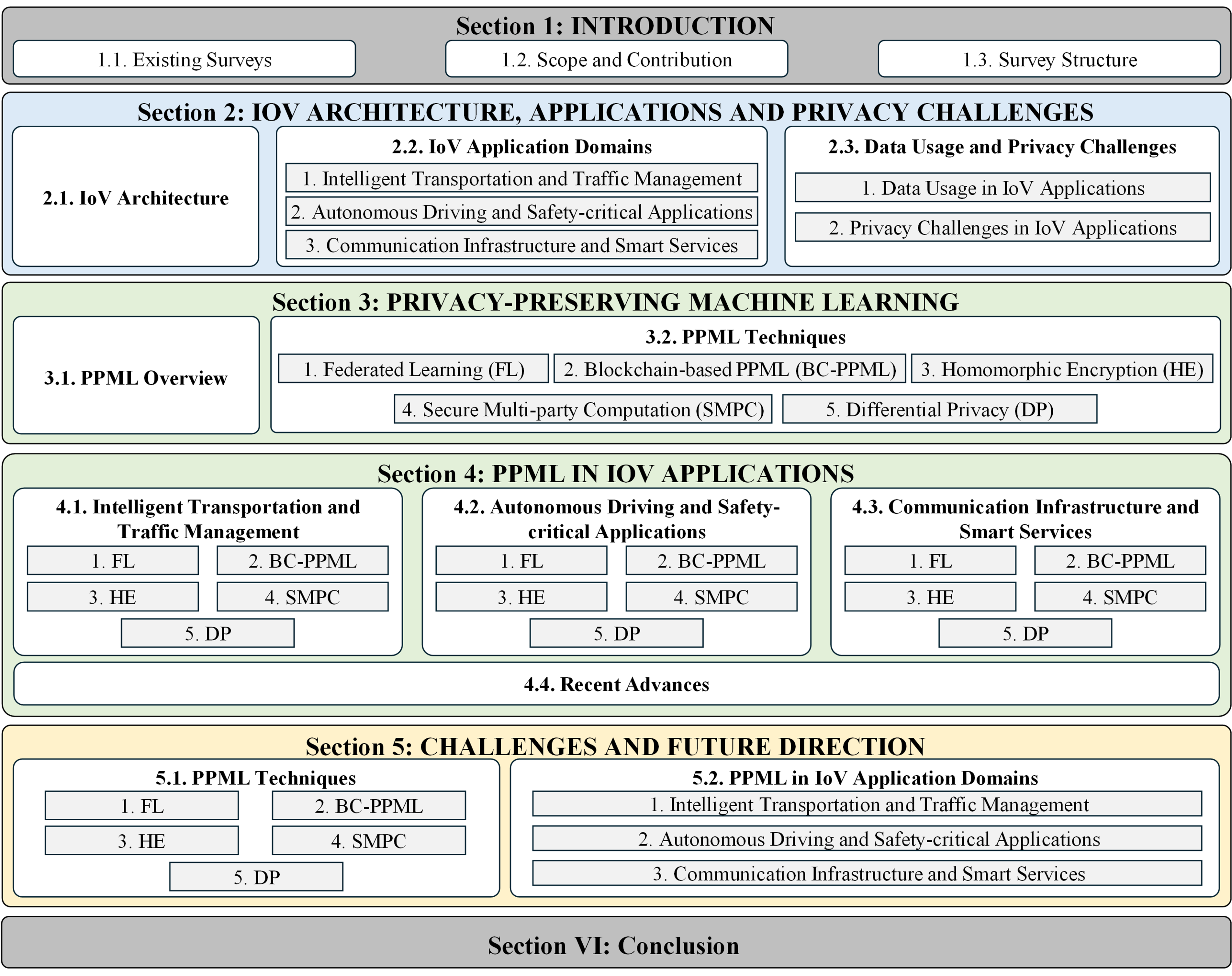}
  \caption{Structural organization of the survey.}\label{fig:roadmap}
\end{figure}

Figure~\ref{fig:roadmap} provides a structural organization of this survey which is as follows. Section~\ref{sec:iov-apps} provides an overview of the IoV ecosystem and the three key application domains within IoV. It also covers the various data type usage within IoV and discusses the associated privacy challenges. Section~\ref{sec:ppml} introduces PPML and the standard techniques, such as FL, BC-PPML, HE, SMPC, and DP. In Section~\ref{sec:ppml-in-iov}, recent advancements in applying these PPML techniques to the IoV application domains are discussed in detail, with focus on their implementation, key features and performance evaluation, along with summary tables. Section~\ref{sec:challenges} identifies recent challenges within PPML techniques and their adoption in IoV applications, offering potential research directions. Section~\ref{sec:conclusion} concludes the survey.

\section{IoV Architecture, Applications and Privacy Challenges}\label{sec:iov-apps}

This section provides the fundamentals of IoV architecture, various applications and their privacy challenges. IoV integrates real-time data acquisition, robust communication networks, intelligent data processing, and application-driven services into a cohesive framework. Various environmental and vehicular data are collected via sensors, actuators and edge devices, which are then transmitted through high-speed, low-latency networks supporting diverse vehicle-to-everything (V2X) interactions \cite{ref36,ref35}. These data streams are processed through ML-based analytics, enabling real-time artificial intelligence (AI) applications for traffic management, autonomous driving or smart infrastructure and services.

\subsection{IoV Architecture}

\begin{figure}[htbp]
  \centering
  \includegraphics[width=0.92\textwidth]{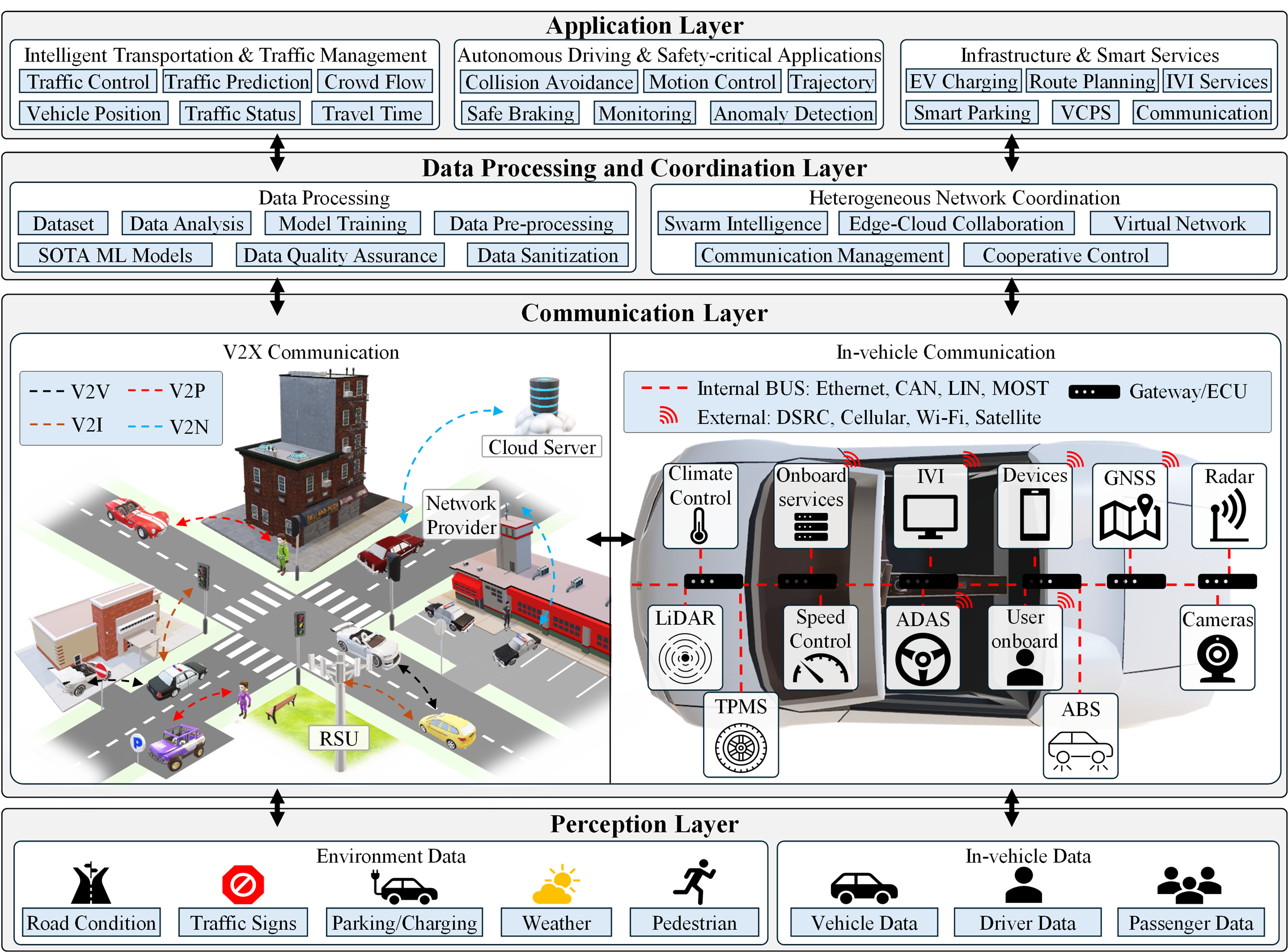}
  \caption{The hierarchical architecture of IoV in four layers.}\label{fig:architecture}
\end{figure}

The IoV architecture consists of multiple layers that work together among vehicles, infrastructure, users, cloud and other entities to enable seamless communication, data processing and applications. Based on the literature review and recent advances in IoV, we have proposed a four-layered IoV architecture illustrated in Figure~\ref{fig:architecture}.

The perception layer in the IoV architecture is responsible for collecting comprehensive data related to vehicles, the environment, and users from various sources, including vehicles, road side units (RSU), smart devices, and other connected data points within the network. The data is collected using different sensors and actuators such as global positioning system (GPS), cameras, ultrasonic sensors, light detection and ranging (LiDAR), accelerometers, radars and magnetometers. This layer is the principal source of all data within the IoV ecosystem \cite{ref36,ref35}.

The communication layer is divided into V2X communication and in-vehicle communication. V2X enables vehicles to exchange information with other vehicles, infrastructure, pedestrians, and the network \cite{ref36}. It comprises vehicle-to-vehicle (V2V), vehicle-to-infrastructure (V2I), vehicle-to-network (V2N), and vehicle-to-pedestrian (V2P) communication. In-vehicle communication systems integrate both external (wireless) and internal (wired) networks to enable seamless data exchange. External network of V2X leverages communication standards to facilitate connectivity with systems such as the global navigation satellite system (GNSS), in-vehicle infotainment (IVI), and advanced driver assistance systems (ADAS). Internal networks enable bus-level communication among electronic control units (ECUs), sensors, controllers, and actuators within the vehicle \cite{ref35}. Collectively, these communication networks ensure reliable and high-speed data transfer across diverse subsystems.

The data processing and coordination layer is responsible for efficient handling of vast amounts of data generated in IoV. This layer integrates advanced technologies including cloud computing for scalable data storage, big data analytics for extracting meaningful insights, and data management systems for ensuring data integrity and accessibility. In some IoV architectures, the data management layer functions as a centralized information management hub, facilitating the storage, processing, and analysis of data received from lower layers, other architectures employ a more distributed approach with edge and cloud layers to optimize latency, bandwidth, and scalability. Additionally, this layer incorporates communication control and management services, enforcing policies such as load balancing and quality of service (QoS) management.

The application layer leverages the vast amounts of data collected and processed by the lower layers to enable intelligent applications and services. This layer utilizes advanced ML and DL algorithms to enable a wide range of intelligent IoV applications. Various AI-based IoV applications are discussed in the following section.

\subsection{IoV Application Domains}

The IoV is a rapidly evolving ecosystem encompassing diverse applications that can be broadly categorized into three key domains: (1) intelligent transportation and traffic management, (2) autonomous driving and safety-critical applications, and (3) communication infrastructure and smart services. Each domain leverages advanced technologies such as AI, the Internet of Things (IoT), V2X communications, and distributed computing with edge-cloud collaboration.

\subsubsection{Intelligent Transportation and Traffic Management}

Intelligent transportation system (ITS) encompasses a wide range of applications designed to enhance safety, efficiency, and sustainability in transportation \cite{ref32}. AI-based computer vision applications, such as object detection, play a pivotal role in ITS by accurately detecting and classifying vehicles, pedestrians, and other road users in real time \cite{ref38}. This technology enables advanced collision prevention systems that predict and mitigate potential accidents, enhancing road safety. Traffic management is another critical aspect of ITS, where AI algorithms analyze real time traffic data to optimize traffic flow, adaptively adjust traffic signals, and provide accurate vehicle positioning utilizing advanced AI algorithms. Additionally, smart public transportation systems utilize AI to optimize routes, predict passenger demand, and improve overall efficiency. Furthermore, AI-driven applications, such as pothole prediction, road damage assessment, and driver misbehavior detection, contribute to enhanced road safety and infrastructure maintenance.

\subsubsection{Autonomous Driving and Safety-critical Applications}

Connected autonomous vehicles (CAVs) integrate advanced communication systems and AI with autonomous driving capabilities to perceive, understand, and navigate complex road environments with minimal human intervention \cite{ref31}. AI algorithms process and interpret vast amounts of data in real time for tasks such as object recognition, trajectory prediction, route planning, and cooperative decision-making \cite{ref41}. CAVs rely on these AI-based perception systems to detect and classify obstacles, pedestrians, and other vehicles by fusing data from multiple sensors and actuators facilitating autonomous driving, while ensuring safety and reliability. Safety-critical applications such as emergency braking or collision avoidance systems are designed to address scenarios requiring high reliability and precision to avoid catastrophic outcomes. Examples include emergency braking systems, collision avoidance mechanisms, and fail-safe operations during sensor or system malfunctions \cite{ref38,ref41}. AI plays a pivotal role in these applications by enabling rapid anomaly detection, redundancy management, and failover protocols to maintain operational safety. Furthermore, reliable communication and coordination between CAVs and other entities through V2X communication are essential for enhancing situational awareness and enabling cooperative decision-making using advanced AI systems.

\subsubsection{Communication Infrastructure and Smart Services}

Communication infrastructure and smart services integrate physical and digital systems to enable intelligent, connected, and automated transportation solutions. A key aspect of smart infrastructure is the charging network for electric vehicles (EVs) \cite{ref47}. AI algorithms optimize charging schedules, predict energy demand, and facilitate smart grid integration, ensuring efficient and sustainable charging operations. Most charging stations (CSs) are strategically located near parking spots to enhance accessibility and convenience. AI-driven parking management systems form another vital component of smart services, leveraging sensors, cameras, and real-time data analytics to guide drivers and CAVs to available parking spots, optimize parking space utilization, and streamline automated payment processes \cite{ref48}. Smart services also offer personalized user experiences, such as infotainment systems to passengers \cite{ref49} or personalized route planning for optimized navigation recommendations. Predictive maintenance is another critical application which leverages AI models to analyze sensor data from vehicles and infrastructure to predict potential failures or optimize maintenance schedules, thereby improving vehicle reliability and assess risk profiles for insurance services \cite{ref50}.

\subsection{Data Usage and Privacy Challenges}\label{sec:data-privacy}

The IoV relies on diverse data types to enable AI-based downstream applications, enhancing transportation efficiency, safety, and user experience. Common data types in IoV and their privacy challenges are discussed in this section.

\subsubsection{Data Usage in IoV Applications}

The IoV leverages various data types to enhance vehicle functionality, user experience, and the integration of smart services. We have categorized the data types that are commonly required for IoV applications. Personal data tracks user travel patterns, frequented locations, and contact information, enabling hands-free connectivity for improved operation and convenience. Biometric data, such as facial recognition and fingerprint scanning, enhances security and personalizes vehicle access and operation \cite{ref51}. Behavioral data analyzes driving habits, including speed, braking, and route choices, optimizing safety, enabling applications such as autonomous driving and predictive maintenance \cite{ref50}. Vehicle operation data continuously monitors metrics like engine status and fuel efficiency, supporting reliability, autonomy and preemptive service. Environmental data on weather, road conditions, and cabin climate improve comfort and driving experience \cite{ref55}. Multimedia and connectivity data store user preferences for navigation, media, and in-vehicle settings while supporting seamless hands-free operations for safety and convenience \cite{ref49}. Road condition and traffic data, collected through vehicle sensors and RSUs, assess traffic flow and pavement quality, providing real-time updates for traffic management, autonomous driving, and road maintenance. Furthermore, IoV requires infrastructure and grid data for EV charging coordination, smart parking, and energy allocation, resource optimization and sustainability.

Among the three key domains, intelligent transportation and traffic management mostly leverages personal, environmental and traffic data to enable advanced navigation, real-time traffic management, hazard warnings and collision prevention \cite{ref38}. Autonomous driving and safety-critical applications depend on all types of personal, behavioral, biometric, operational, road and environmental data to support autonomous driving, secure access, adaptive safety systems, advanced navigation and route optimization \cite{ref38,ref41}. Meanwhile, communication infrastructure and smart services mostly utilizes multimedia, infrastructural, environmental, operational and energy data to facilitate smart parking \cite{ref48}, charging coordination, predictive maintenance \cite{ref55}, and personalized infotainment \cite{ref49}.

\subsubsection{Privacy Challenges in IoV Applications}

The adoption of AI in IoV applications raises privacy concerns due to the large amounts of data collected, transmitted and stored from vehicles, passengers, infrastructure, and pedestrians \cite{ref4,ref6}. A primary challenge is the exposure of Personally Identifiable Information (PII), including location, driving patterns, and biometric data. Vehicle-specific data, such as make, model, and usage, can also increase the risk of targeted attacks \cite{ref22,ref23,ref24}. Data security is another critical challenge, as the transmission and storage of sensitive data create vulnerabilities that may lead to breaches, exposing both individual and aggregate information \cite{ref11,ref12,ref20}. Furthermore, unauthorized access to models can compromise privacy, as AI algorithms may inadvertently expose personal details through inference, even when data is anonymized.

In the ML pipeline of an IoV application, privacy risks can arise during the training and inference phases to compromise the confidentiality and integrity of the system \cite{ref8,ref9,ref10}. In the training phase, data poisoning attacks involve adversaries manipulating training data to compromise model integrity or insert backdoors, a significant risk in IoV where data quality is crucial for safety. Property inference attacks allow malicious participants to extract statistical properties about other models/users private training data, potentially exposing vehicle patterns or user behaviors. Byzantine attacks (common in distributed learning scenarios) involve malicious vehicles or data points injecting false gradients during training, destabilizing the model impacting its efficiency and reliability.

During the inference phase, privacy risks are amplified by attacks like membership inference, model inversion, and model extraction \cite{ref8}. Membership inference attacks enable adversaries to determine whether specific vehicle data was used in training, potentially revealing sensitive patterns. Model inversion attacks aim to reconstruct sensitive training data from model parameters, risking exposure to personal vehicle information, such as location histories. Model extraction attacks involve attempts to steal the functionality of the model through carefully crafted queries, potentially compromising proprietary vehicle behavior models \cite{ref8,ref24}.

The involvement of third-party services for data collection, training, or model deployment further exacerbates privacy concerns. Third parties can mishandle data, expose sensitive information, or become targets of separate breaches, jeopardizing the entire ecosystem. If adversaries access third-party data or training pipelines, they could launch targeted attacks that affect all connected vehicles using the compromised system. Data breaches within third-party services could lead to the loss of proprietary vehicle models or private data, compromising user privacy and vehicle security on a large scale \cite{ref11,ref12,ref20}. Additionally, man-in-the-middle or other interception attacks on communication channels between vehicles and third-party infrastructures can intercept or alter data, threatening privacy and security of IoV applications.

\section{Privacy-preserving Machine Learning}\label{sec:ppml}

Privacy-preserving machine learning (PPML) integrates various privacy-preserving techniques into ML pipelines to mitigate privacy risks in collaborative learning scenarios. As IoV systems handle vast amounts of sensitive and private data, the need for PPML solutions has become increasingly critical on both the user and server sides.

\subsection{PPML Overview}

The focus of this study is on three distinct types of PPML approaches integrated into IoV application domains. Firstly, the architecture-based PPML approach that represent distributed architectural solutions of federated learning (FL) and blockchain-based PPML (BC-PPML). FL enables privacy-preserving distributed learning, while BC-PPML adds an additional layer of transparency and accountability to the ML process, ensuring secure model updates and verification. Secondly, the data processing-based PPML methods that include homomorphic encryption (HE) and garbled circuits, where HE enables computation on encrypted data and garbled circuits facilitate secure multi-party computation (SMPC), protecting intermediate calculations during training. Finally, the data publishing approach of differential privacy (DP) which introduces statistical noise to protect individual privacy while maintaining useful aggregate information for training.

PPML operates across three main phases: model generation, model training, and model serving \cite{ref8}. During model generation, privacy protection is implemented in the initial development and architecture design stages. Model training incorporates privacy-preserving techniques throughout the learning process, while model serving ensures privacy during deployment and inference stages, protecting both user queries and model outputs. The privacy guarantees in PPML can be categorized into two main categories \cite{ref8}. Object-oriented guarantees focus on protecting specific components, including data-oriented protection for raw training data and user inputs, and model-oriented protection for model parameters and architecture. Pipeline-oriented guarantees are generic end-to-end privacy protection across the entire ML workflow. Table~\ref{tbl:ppml-comparison} presents a comparative analysis of the PPML techniques and their performance in terms of privacy, utility, and efficiency.

\begin{table}[htbp]
\centering
\scriptsize\normalfont
\setlength{\tabcolsep}{4pt}
\caption{Comparison of various PPML techniques
(\cmark\cmark\ strongly favorable, \cmark\ favorable, \pmark\ partial/mixed,
\xmark\ unfavorable, \xmark\xmark\ strongly unfavorable, -- NA).}
\label{tbl:ppml-comparison}
\setlength{\tabcolsep}{2pt}
\begin{tabular}{
@{}
>{\raggedright\arraybackslash}p{0.09\linewidth}
>{\raggedright\arraybackslash}p{0.315\linewidth}
>{\centering\arraybackslash}p{0.11\linewidth}
>{\centering\arraybackslash}p{0.12\linewidth}
>{\centering\arraybackslash}p{0.11\linewidth}
>{\centering\arraybackslash}p{0.11\linewidth}
>{\centering\arraybackslash}p{0.11\linewidth}
@{}
}
\textbf{Category} &
\textbf{Criterion} &
\textbf{FL \cite{ref19,ref20,ref21}} &
\textbf{BC \cite{ref15, ref29, ref30}} &
\textbf{HE \cite{ref17, ref26}} &
\textbf{SMPC \cite{ref12, ref18}} &
\textbf{DP \cite{ref16, ref27}} \\
\midrule
\multirow{7}{*}{\centering{Privacy}}
&
No raw data/plaintext exposed to other parties
& \cmark
& \pmark$^{a}$
& \cmark\cmark
& \cmark\cmark
& \cmark
\\
&
Formal/quantifiable privacy guarantee
& \xmark
& --
& \cmark\cmark
& \cmark\cmark
& \cmark\cmark
\\
&
Resistant to poisoning or adversarial participants
& \xmark\xmark
& \cmark
& --
& \pmark$^{b}$
& --
\\
&
Provides auditability / verifiable computation
& --
& \cmark\cmark
& --
& --
& --
\\
&
Tamper-evident record of operations
& --
& \cmark\cmark
& --
& --
& --
\\
&
Guarantee strength is configurable/tunable
& --
& --
& \cmark
& --
& \cmark
\\
&
Free of a documented privacy-erosion risk
& \xmark
& \pmark$^{c}$
& --
& --
& \xmark
\\
\midrule
\multirow{5}{*}{\centering{Utility}}
&
Minimal accuracy loss vs.\ centralized training
& \pmark
& --$^{d}$
& --
& \cmark
& \xmark\xmark
\\
&
Supports complex/non-linear ML operations natively
& \cmark
& --$^{d}$
& \xmark\xmark
& \xmark
& \cmark
\\
&
Produces exact (non-approximate) results
& --
& --
& --
& \cmark
& \xmark\xmark
\\
&
Practical for full model training (not just inference)
& \cmark\cmark
& --
& \xmark\xmark
& --
& \cmark
\\
&
Compatible with standard ML pipelines/optimizers
& \cmark
& --
& \xmark
& --
& \cmark\cmark
\\
\midrule
\multirow{6}{*}{\centering{Efficiency}}
&
Low computational overhead
& --
& \xmark
& \xmark\xmark
& \xmark
& \cmark\cmark
\\
&
Low communication overhead
& \xmark\xmark
& --
& --
& \xmark
& \cmark\cmark
\\
&
Low storage overhead
& --
& \xmark
& \xmark
& --
& \cmark\cmark
\\
&
Scales well to large numbers of participants/nodes
& \xmark
& \xmark
& --
& \xmark\xmark
& \cmark\cmark
\\
&
Efficiency improves with implementation choice
& --
& \cmark$^{e}$
& --
& \cmark$^{e}$
& --
\\
&
Low key/protocol management complexity
& --
& --
& \xmark
& \xmark
& --
\\
\bottomrule
\multicolumn{7}{@{}p{\linewidth}@{}}{
\scriptsize

$^{a}$BC-PPML alone does not guarantee data confidentiality; it is typically paired with HE/SMPC/DP for this purpose.
$^{b}$SMPC's threshold security tolerates a bounded number of corrupted parties, not unlimited adversarial participation.
$^{c}$This risk is explicitly conditional on implementation quality, unlike FL's and DP's inherent risks.
$^{d}$BC-PPML's utility gain is verifiability/provenance, not ML accuracy, so it is not directly comparable to the other techniques on this row.
$^{e}$Efficiency gains are conditional on implementation choice (BC-PPML: consensus protocol; SMPC: optimized 3-/4-party protocols).

} \\
\end{tabular}
\end{table}

\subsection{PPML Techniques}\label{sec:ppml-techniques}

This section discusses the fundamentals of the PPML techniques covered in the study, including their mathematical foundations and types. It also discusses how these techniques are applied in the context of ML-based IoV applications.

\subsubsection{Federated Learning}

\begin{figure}[htbp]
  \centering
  \includegraphics[width=0.92\textwidth]{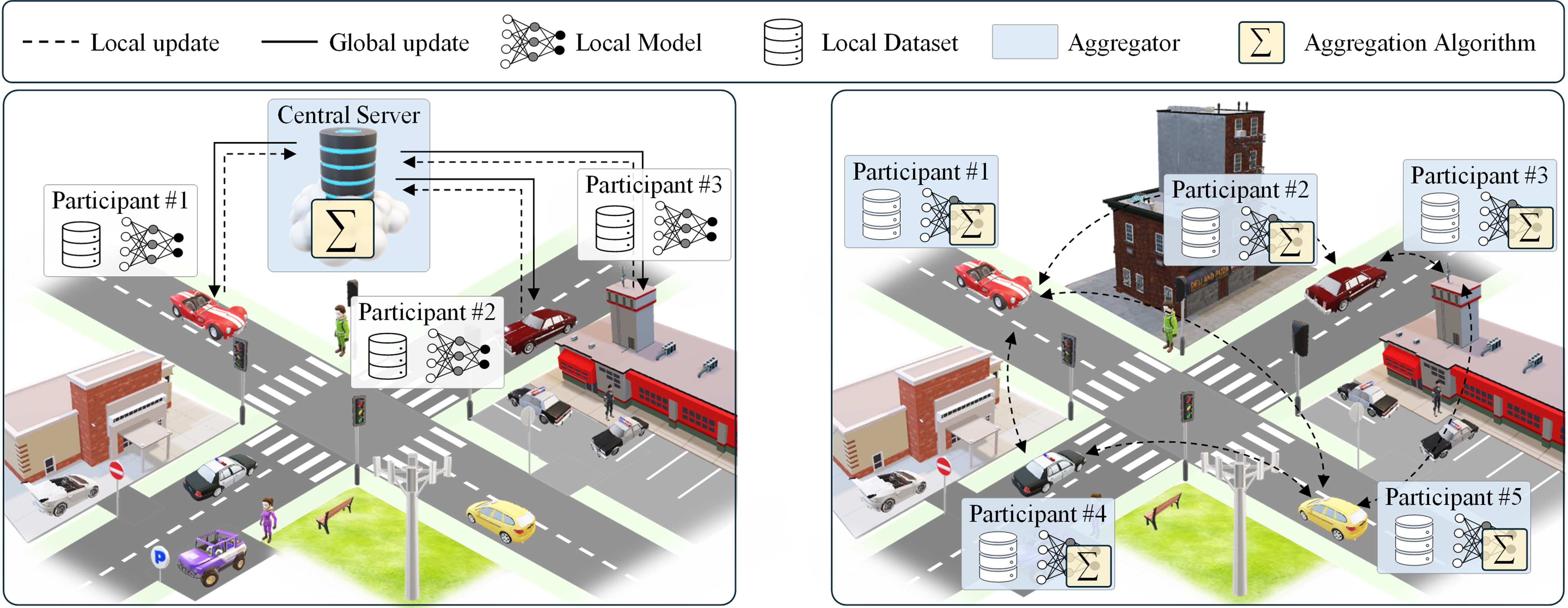}
  \caption{Basic illustration of centralized FL (left) and decentralized FL (right) structure in IoV.}\label{fig:fltypes}
\end{figure}

FL is a PPML technique that enables training models on distributed datasets without directly sharing raw data. This method not only enhances data privacy but also addresses challenges associated with data heterogeneity and communication efficiency \cite{ref8}. As illustrated in Figure~\ref{fig:fltypes}, FL can be categorized into two main types, centralized FL (CFL) and decentralized FL (DFL) \cite{ref20}. In CFL, participants (also known as clients or nodes) train models locally using their data and send the updates (usually model gradients or parameters) to the central server. The server refines the model by aggregating the updates using various algorithms and then sends back the updated global model to clients for further training, inference, or deployment. Two popular aggregation algorithms are federated averaging (FedAvg) and federated distillation \cite{ref20}. FedAvg performs weighted averaging of client updates for data aggregation at the server side using:

\begin{equation}\label{eq:fedavg}
\theta^{(t+1)} = \sum_{v \in V} \frac{n_{v}^{(t)}}{n^{(t)}}\theta_{v}^{(t+1)}
\end{equation}

where, $\theta_{v}^{(t+1)}$ are the updated parameters from vehicle $v$, and $n_{v}^{(t)}$ is the size of its local dataset. On the client side, the objective function for the local training of the $v$th vehicle can be expressed as:

\begin{equation}\label{eq:client-obj}
f_{v}(\theta_{v}) = E_{z_{v}\sim D_{v}}\left[l(\theta_{v},z_{v})\right]
\end{equation}

where $\theta_{v}$ are the model parameters, $z_{v}$ is a data sample from the local dataset $D_{v}$, and $l(\theta_{v},z_{v})$ is the loss function.

FedAvg falls short in real-world heterogeneous data environments, since a single global model may not suit the individual vehicles, and frequent local updates can result in divergence from the global objective. Knowledge distillation overcomes the challenges of model heterogeneity and data heterogeneity by allowing different clients to use varied model architectures (by exchanging logits instead of model parameters), and mitigating non-independent and identically distributed (non-IID) issues in data respectively. In knowledge distillation the knowledge (typically in the form of logits) is transferred from a complex teacher model to simpler student models. In this process, each client trains its local model with its data and generates logit outputs, which are shared with a central server. The server aggregates these logits to form a global logit, representing the distilled knowledge from the teacher model. Clients then use this global logit to train their local models, aligning the student's outputs with the teacher's soft targets. The student is trained using linear combination of loss functions:

\begin{equation}\label{eq:kd-loss}
L = (1-\lambda)L_{CE}(q_{s},y) + \lambda L_{KL}\left(q_{s}^{\tau},q_{T}^{\tau}\right)
\end{equation}

where $L_{CE}$ is cross-entropy loss between the student's predictions $q_{s}$ and true labels $y$. $L_{KL}$ is Kullback-Leibler divergence between the student's soft targets $q_{s}^{\tau}$ and the teacher's soft targets $q_{T}^{\tau}$, $\lambda$ is a hyperparameter that balances the two loss components. $\tau$ is a temperature parameter to smooth the probability distribution produced by SoftMax.

While CFL is effective in various scenarios, it has limitations such as a single point of failure, bottlenecks at the server, and various privacy risks associated with centralized data aggregation \cite{ref20}. DFL addresses these limitations by eliminating the need for a central server. Instead, each client trains its local model independently and exchanges model updates with its peers (participating nodes). This enhances fault tolerance, robustness against single points of failure and reduces trust dependencies on a central entity. In DFL, achieving consensus among participating nodes is crucial to ensure coherent model updates and reflect the contributions of all \cite{ref20}. A common approach to achieve consensus involves using a mixing matrix $w_{ij}(t)$, where each element $w_{ij}$ represents the weight of communication between node $i$ and node $j$. The update rule for each peer's model is:

\begin{equation}\label{eq:dfl-consensus}
\theta_{i}(t+1) = \sum_{j=1}^{n} w_{ij}(t)\theta_{j}(t)
\end{equation}

The number of nodes is given by $n$ and $\theta(t)$ represents the state of a node at time $t$. The matrix ensures that each peer's update considers information from its neighbors, weighted by the strength of their connection. This balances contributions from different nodes and achieves consensus without a central coordinator. The communication among peers can be organized based on different network topologies, such as fully connected networks, partially connected networks, or clustered networks. The choice of topology affects the overall privacy, robustness, flexibility, fault tolerance, and communication costs of the network \cite{ref20}.

\subsubsection{Blockchain-based PPML}

The decentralized and cryptographic nature of BC technology addresses several privacy concerns inherent in traditional, centralized ML architectures. BC provides a decentralized ledger that records transactions securely and immutably. This ledger is maintained across multiple nodes, ensuring there is no single point of failure and no single entity has control over the entire dataset, thus preserving privacy. BC also employs various cryptographic techniques to secure data. Each transaction or data entry on the BC is hashed and linked to previous entries, forming an immutable chain. This ensures that once data is recorded, it cannot be altered without consensus from the network, providing robustness against tampering and unauthorized modifications \cite{ref30}.

BC consists of several key layers. The data layer stores transaction data using cryptographic methods for data integrity and confidentiality. The network layer enables peer-to-peer communication for secure data broadcasting and authentication. The consensus layer employs lightweight algorithms such as practical byzantine fault tolerance (PBFT) or directed acyclic graphs (DAG) to achieve agreement on transaction validity without energy-intensive mining processes \cite{ref30}. Lastly, the application layer supports various IoV applications by utilizing smart contracts to automate processes, prevent unauthorized interventions and ensure secure operations.

BC technology can be combined with other privacy-preserving techniques to further enhance privacy. BC supports pseudonymity by allowing users to interact with the system through pseudonymous addresses rather than real identities. This feature is crucial for PPML, as it prevents the exposure of participants' identities during data transactions and model training processes. Advanced cryptographic techniques such as zero-knowledge proofs (ZKP) and HE integrated into BC systems also significantly enhance data privacy. ZKPs allow one party to prove to another that a statement is true without revealing any information beyond the validity of the statement itself. This can be particularly useful in ML for verifying computations or model parameters without exposing underlying data. Furthermore, BC can enhance FL by providing a secure and decentralized framework for aggregating model updates from different parties. It can also introduce incentive mechanisms to encourage honest participation and fair behavior \cite{ref30}.

\subsubsection{Homomorphic Encryption}

PPML utilizes HE to maintain data confidentiality during the training and inference phases of ML pipeline. HE allows computations to be performed on encrypted data without needing to decrypt it, preserving privacy of sensitive user data. $HE = (KeyGen, Enc, Dec, Eval)$, is composed of four probabilistic polynomial-time algorithms and they are defined as follows. Key generation ($HE_{KeyGen}$) takes a security parameter $\lambda$ as input and outputs a public key ($pk$), a secret key ($sk$), and an evaluation key ($ek$), represented as $(pk,sk,ek) \leftarrow HE_{KeyGen}(\lambda)$. Encryption ($HE_{Enc}$) involves taking the public key $pk$ and a plaintext message $m$ as inputs to produce a ciphertext $c$, expressed as $c \leftarrow HE_{Enc}(pk,m)$. Decryption ($HE_{Dec}$) requires the secret key $sk$ and the ciphertext $c$ as inputs to output the decrypted message $m^{*}$, denoted as $m^{*} \leftarrow HE_{Dec}(sk,c)$. Evaluation ($HE_{Eval}$) takes the evaluation key $ek$, a function $f$, and a series of ciphertexts $c_{0},\ldots,c_{l-1}$ as inputs. Here, each ciphertext $c_{i}$ corresponds to a plaintext $m_{i}$ for $i = 0,\ldots,l-1$, where $l$ is the number of ciphertexts. The evaluation outputs a final ciphertext $c_{fin}$, expressed as $c_{fin} \leftarrow HE_{Eval}(ek,f,c_{0},\ldots,c_{l-1})$ such that $HE_{Dec}(sk,c_{fin}) = f(m_{0},\ldots,m_{l-1})$. The function $f$ represents an operational circuit over the plaintext space.

HE can be broadly categorized into three types: partial HE, somewhat (or strong) homomorphic, and Fully HE (FHE) \cite{ref64}. Partially HE supports unlimited operations of a single type (such as addition or multiplication). For instance, the Rivest-Shamir-Adleman (RSA) cryptosystem allows multiplication and the Paillier cryptosystem supports addition on encrypted data without decryption. The Boneh--Goh--Nissim (BGN) cryptosystem is well-known for supporting both addition and multiplication but only allows a limited number of operations, making it somewhat HE. The first FHE scheme was proposed using lattice-based cryptography, which introduced the concept of bootstrapping to refresh ciphertexts and manage noise accumulation \cite{ref64}. This has led to schemes based on lattice theory like learning with errors (LWE) and ring-LWE, as well as those based on approximating the greatest common divisor \cite{ref64}.

FHE is a powerful tool for PPML because it allows for arbitrary computations on encrypted data, ensuring that both training and inference processes can be conducted securely. For instance, consider a simple linear model represented by the equation $y = w^{t}x + b$ where $w$ is the weight vector, $x$ is the input feature vector, and $b$ is the bias term. In this context, FHE enables each component to be encrypted: the input as $E(x)$, the weights as $E(w)$, and the bias as $E(b)$. The model can then compute the encrypted output as $E(y) = E(w^{t}x+b) = E(w^{t}x) + E(b)$. This preserves data privacy throughout the ML computation process. Despite various advances, high computation and complexity associated with HE still persist when integrating in PPML \cite{ref64}.

\begin{figure}[t]
  \centering
  \includegraphics[width=\columnwidth]{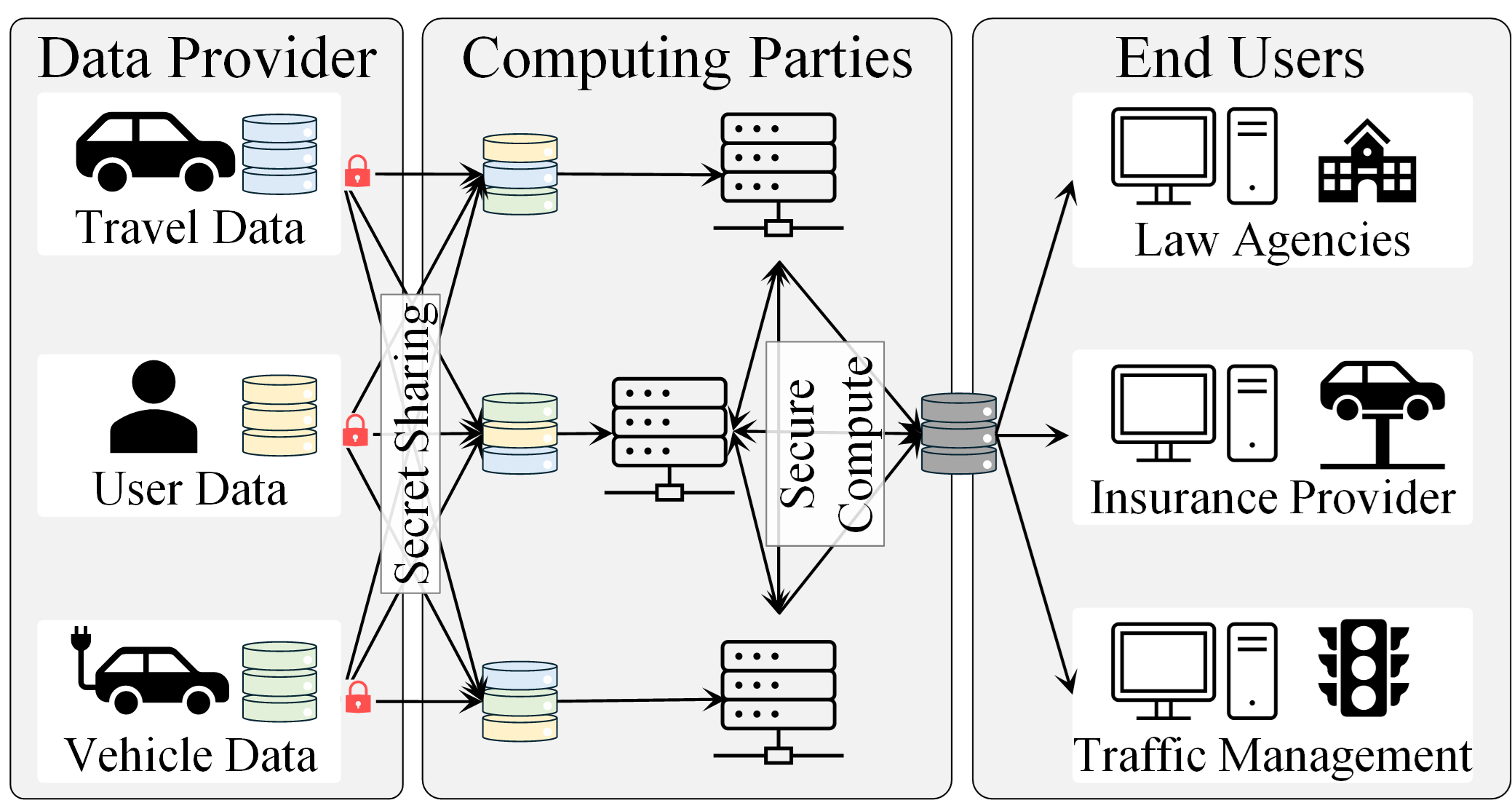}
  \caption{General illustration of SMPC architecture.}\label{fig:smpc}
\end{figure}

\subsubsection{Secure Multi-party Computing}

The goal of SMPC is to create a secure protocol that allows multiple participants $P_{i}$, where $i = 1,\ldots,m$ to jointly compute a function $f(x_{1},\ldots,x_{m}) = (y_{1},\ldots,y_{m})$ based on their private inputs $x_{i}$ \cite{ref13}. Each participant $P_{i}$ should receive only their respective output $y_{i}$, without gaining extra knowledge. This ensures input privacy and correctness, where no information about individual participants can be inferred beyond what the output reveals, and even if some parties collude or deviate from the protocol, they cannot impose an incorrect result on honest parties \cite{ref8}. SMPC can leverage several cryptographic techniques such as secret sharing and secure computation to further enhance privacy \cite{ref13}. In secret sharing data is divided into shares and distributed among participants, ensuring that only specific combinations of these shares can reconstruct the original data. This ensures that no single party has access to the complete dataset, maintaining privacy across the board. In secure schemes (like HE), computations are performed on encrypted data enabling privacy throughout the lifecycle \cite{ref13}.

In the context of PPML, SMPC allows for collaborative model training where multiple parties can jointly train using their private datasets as shown in Figure~\ref{fig:smpc}. SMPC also supports secure inference, enabling parties to perform model inference while ensuring that neither the model owner nor the input provider learns about information more than necessary. Furthermore, SMPC can be applied to feature selection, allowing parties to collaboratively determine important features for a model without exposing raw feature values, enhancing model accuracy while preserving privacy.

\subsubsection{Differential Privacy}

\begin{figure}[htbp]
  \centering
  \includegraphics[width=0.92\textwidth]{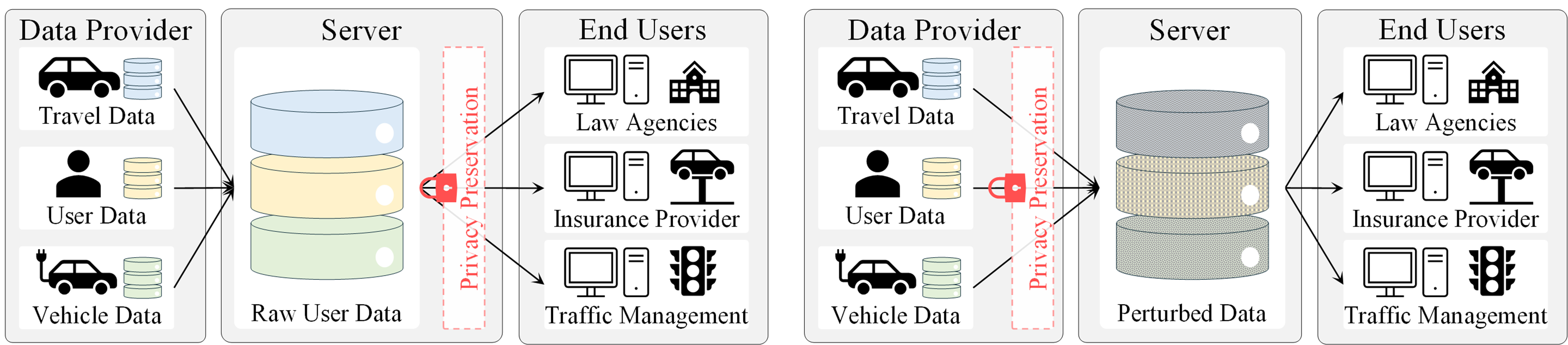}
  \caption{Illustration of centralized DP (left) and local DP (right) for IoV applications.}\label{fig:dp}
\end{figure}

DP provides a mathematical guarantee that individual data points cannot be inferred from the output of a model \cite{ref16,ref27}. It achieves this by introducing controlled randomness into the data analysis process, ensuring that the inclusion or exclusion of any single data point does not significantly affect the outcome. In distributed ML, DP helps protect user data by adding noise to the model updates before they are aggregated. DP operates on the principle that a privacy-preserving mechanism should produce outputs that are statistically indistinguishable whether or not any single client data is included \cite{ref16,ref27}. A mechanism $M$ satisfies $\varepsilon$-DP if, for any two neighboring datasets $d$ and $d'$ and for any possible output subset $S$, the probability $\Pr$ that $M$ produces an output in $S$ is nearly the same for both datasets:

\begin{equation}\label{eq:dp-def}
\Pr\left[M(d) \in S\right] \leq e^{\varepsilon}\Pr\left[M(d') \in S\right]
\end{equation}

Here, $\varepsilon$ is a non-negative parameter that quantifies the privacy loss. Smaller values of $\varepsilon$ indicate stronger privacy.

The noise or controlled randomness in DP is often drawn from a Laplacian distribution, which is determined by the sensitivity of the function being computed. Sensitivity, denoted as $\Delta_{p} Q$, measures how much the output can change when a client's data is added or removed:

\begin{equation}\label{eq:sensitivity}
\Delta_{p} Q = \max_{d,d'} \left\|Q(d) - Q(d')\right\|_{p}
\end{equation}

where the maximum is taken over all neighboring datasets $d$ and $d'$; $p=1$ is used for the Laplace mechanism and $p=2$ for the Gaussian mechanism.

Noise drawn from a Laplace distribution is typically added to the function's output to obscure individual contributions:

\begin{equation}\label{eq:laplace}
M(d) = Q(d) + \mathrm{Laplace}\left(0,\frac{\Delta_{1} Q}{\varepsilon}\right)
\end{equation}

here, $Q(d)$ represents the true query result, and Laplace noise ensures that individual contributions are obscured. A relaxed version known as $(\varepsilon,\delta)$-DP allows for a small probability $\delta$ that the strict privacy guarantee might be violated:

\begin{equation}\label{eq:relaxed-dp}
\Pr\left[M(d) \in S\right] \leq e^{\varepsilon}\Pr\left[M(d') \in S\right] + \delta
\end{equation}

This relaxation provides more flexibility in designing mechanisms with privacy gurantee tradeoff. The Gaussian mechanism, which scales noise according to the L2 norm, is commonly used:

\begin{equation}\label{eq:gaussian-mech}
M(d) = Q(d) + N\left(0,\frac{2\ln(1.25/\delta)\left(\Delta_{2}Q\right)^{2}}{\varepsilon^{2}}\right)
\end{equation}

where $N(0,2\ln(1.25/\delta)(\Delta_{2}Q)^{2}/\varepsilon^{2})$ represents Gaussian noise with variance calibrated to satisfy $(\varepsilon,\delta)$-DP, scaled according to sensitivity. This calibration holds for $\varepsilon \in (0,1)$; larger values of $\varepsilon$ require a different noise calibration.

As illustrated in Figure~\ref{fig:dp}, DP can be broadly categorized into two main approaches, centralized and local DP. These techniques can be implemented through various mechanisms in DL, including adding random noise to input samples, gradients, or functions during the training process \cite{ref16,ref27}. Implementing DP involves balancing privacy and utility. More noise generally implies better privacy but degrades model accuracy. Additionally, managing cumulative privacy loss over multiple queries (the privacy budget) is crucial to maintaining robust privacy guarantees over time.

\FloatBarrier

\section{PPML in IoV Applications}\label{sec:ppml-in-iov}

The following subsections review advancements in integrating PPML techniques into IoV applications. The literature is categorized into the three key IoV domains: (1) intelligent transportation and traffic management, (2) autonomous driving and safety-critical applications, and (3) communication infrastructure and smart services. Within these domains, studies are further classified based on the PPML techniques.

\subsection{Intelligent Transportation and Traffic Management}\label{sec:its}

For intelligent transportation and traffic management, real-time data aggregation and analysis are employed to optimize traffic flow, reduce congestion, and enhance urban mobility. The integration of various PPML techniques in this domain enables decentralized, secure data sharing and computation while maintaining privacy. We review the recent advances in this domain and the technical summary of relevant literature is given in Table~\ref{tbl:its-summary}.

{\scriptsize
\setlength{\tabcolsep}{3pt}
\begin{longtable}{@{}>{\raggedright\arraybackslash}p{0.055\linewidth}>{\raggedright\arraybackslash}p{0.095\linewidth}>{\raggedright\arraybackslash}p{0.33\linewidth}>{\raggedright\arraybackslash}p{0.13\linewidth}>{\raggedright\arraybackslash}p{0.33\linewidth}@{}}
\caption{Literature adopting PPML in intelligent transportation and traffic management systems.}\label{tbl:its-summary}\\
\toprule
\textbf{Ref.} & \textbf{Tech. (Trust)} & \textbf{Application (Dataset)} & \textbf{Privacy target} & \textbf{Notable Outcome} \\
\midrule
\endfirsthead
\caption[]{Literature adopting PPML in intelligent transportation and traffic management systems. (continued)}\\
\toprule
\textbf{Ref.} & \textbf{Tech. (Trust)} & \textbf{Application (Dataset)} & \textbf{Privacy target} & \textbf{Notable Outcome} \\
\midrule
\endhead
\midrule
\multicolumn{5}{r}{\textit{Continued on next page}} \\
\endfoot
\bottomrule
\multicolumn{5}{@{}p{\linewidth}@{}}{\scriptsize Tech.: primary technique first, then any combined techniques (hybrid PPML). Trust assumption in brackets: C central aggregator, S semi-trusted edge/fog/RSU/cloud, N non-colluding servers, D decentralized (no trusted party).} \\
\endlastfoot
\cite{ref71} & FL~(C) & Traffic flow prediction (PeMS04/08) & Model & High accuracy; Low overhead with low time cost \\
\cite{ref72} & FL+\allowbreak{}DP~(C) & Traffic speed forecasting (PeMSD7) & Network/\allowbreak{}model & robust to time-series fluctuations \\
\cite{ref73} & FL+\allowbreak{}HE~(C) & Traffic flow prediction (TaxiNYC/TaxiBJ) & Spatial-temporal & Improved accuracy; edge-optimized \\
\cite{ref74} & FL~(C) & Traffic flow prediction (PeMS) & Parameter & 0.76 km/h error; reduced overhead \\
\cite{ref75} & FL~(C) & Graph-based traffic forecasting (PeMSD4/PeMSD7) & Model & Satisfactory accuracy; lower comm. overhead \\
\cite{ref76} & FL~(C) & Traffic state estimation (England Freeway) & State & Good state accuracy; resource-optimized \\
\cite{ref77} & FL+\allowbreak{}DP~(C) & Crowd flow prediction (Generated) & Gradient & Improved accuracy; reduced communication \\
\cite{ref78} & FL~(C) & Mobility forecasting (New York City/Yelp) & Data partition & Cross-org security; 4--12\% accuracy improvement \\
\cite{ref83} & FL~(C) & Adaptive signal control (Generated) & Model & Enhanced control security; Better optimality \\
\cite{ref84} & FL~(C) & Multi-intersection signal control (Cityflow, sim.) & Model integrity & Superior performance; Resource-optimized \\
\cite{ref85} & FL~(S) & Edge caching for traffic control (MNIST/MovieLens) & Edge/\allowbreak{}Vehicle & High hit rate; low latency \\
\cite{ref86} & FL~(C) & Congestion monitoring (LA/Washington Road) & Data & 85\% detection rate; 0.047 s latency \\
\cite{ref87} & FL~(C) & Cooperative vehicle positioning (Didi Chengdu) & Position/\allowbreak{}trajectory & Improved convergence; Low overhead \\
\cite{ref90} & FL~(C) & GPS-free vehicle tracking (Didi Beijing/Shanghai) & Location/\allowbreak{}sensor & Lowest MAE; 2.25 MB storage \\
\cite{ref91} & FL~(C) & Travel mode identification (GeoLife) & Trajectory/\allowbreak{}model & 90\% accuracy with 50\% labeled data \\
\cite{ref283} & FL~(C) & Pedestrian trajectory prediction (ETH/UCY/SDD) & Data & Lightweight; Distributed scene \\
\cite{ref279} & FL~(C) & Traffic flow/speed prediction (METR-LA/Luzern) & Data & Reduced RMSE by 6.7\%; Low space complexity \\
\cite{ref281} & FL~(C) & Traffic/speed/occupancy pred. (UTD19 + 3 datasets) & Data & Outperforms 24 baselines; Handles heterogeneity \\
\cite{ref282} & FL~(C) & Trust-aware traffic forecasting (METR-LA) & Data & MAE 2.369; Stable convergence \\
\cmidrule{1-5}
\cite{ref69} & BC~(D) & ITS intrusion detection (ToN-IoT/CICIDS-2017) & Data & Prevent inference attack; Over 98\% accuracy \\
\cite{ref93} & BC~(D) & Traffic estimation (Generated) & User & High accuracy; High participation \\
\cite{ref94} & BC+\allowbreak{}FL+\allowbreak{}DP~(D) & Traffic prediction (PeMS) & Vehicle & Robust to poisoning; High computation \\
\cite{ref95} & BC~(D) & Real-time traffic prediction (DelDOT) & Real-time & High throughput; Low latency \\
\cite{ref96} & BC+\allowbreak{}FL+\allowbreak{}HE~(D) & Traffic prediction (DelDOT) & Parameter & Improved acc. (controlled group; costly) \\
\cite{ref98} & BC+\allowbreak{}FL+\allowbreak{}HE+\allowbreak{}DP~(D) & Fair FL framework (MNIST/SVHN) & Model & High accuracy; low utility cost \\
\cite{ref99} & BC~(D) & Trust management (Foursquare) & Reputation & Fair trust evaluation; high throughput \\
\cite{ref100} & BC~(D) & Task assignment (MNIST) & Worker/\allowbreak{}training & Optimal matching; resource-optimized \\
\cite{ref101} & BC+\allowbreak{}FL~(D) & Mobile edge FL (MNIST) & Model & High accuracy; reduced delay \\
\cite{ref270} & BC~(D) & Data sharing in C-V2X (Sim. + real highway data) & Data & 90\% less on-chain cost; 31.7\% lower delay \\
\cite{ref271} & BC+\allowbreak{}HE~(D) & Emission-aware speed advisory (Sim. ICEV/EV fleet) & Data & Minimizes emissions; under 300 ms, over 100 TPS \\
\cite{ref277} & BC+\allowbreak{}FL~(D) & Trajectory prediction (nuScenes) & Trajectory & Best MinADE on benchmark (1.14--1.39 m) \\
\cite{ref278} & BC+\allowbreak{}FL~(D) & Traffic sign recognition (GTSRB) & Model & Robust under label-flip attack (>92\% acc.) and unavailable base-station \\
\cite{ref269} & BC~(D) & Accident forensics (Hyperledger/Caliper benchmark) & Accident records & Secure access control; collision-resistant; 0.02--0.12 s \\
\cite{ref273} & BC~(D) & Crowdsourcing task matching (MTurk HITs) & Worker/\allowbreak{}requester & 12--23\% faster indexing; 6.8x faster matching \\
\cite{ref272} & BC+\allowbreak{}HE~(S) & Task offloading (Simulated VECON) & Task data & Lowest joint delay-energy-privacy cost \\
\cmidrule{1-5}
\cite{ref102} & HE+\allowbreak{}FL~(C) & Traffic flow prediction (PeMS) & Data/\allowbreak{}Model & Training loss below 0.003 \\
\cite{ref103} & HE+\allowbreak{}DP~(S) & Vehicular fog navigation (Generated) & Parameter & Flexible participation; improved computation \\
\cite{ref104} & HE+\allowbreak{}BC~(D) & Decentralized VANET security (Generated) & Model credibility & Attack prevention; Improved throughput \\
\cite{ref280} & HE+\allowbreak{}SMPC+\allowbreak{}DP~(C) & Secure parameter aggregation (Traffic sign + DALL-E) & Model & Protect against reverse/membership inference; Lightweight \\
\cite{ref105} & HE~(S) & IoV crowd sensing (SUMO-generated) & Data & Data/access protection; High throughput in dense traffic \\
\cite{ref106} & HE~(S) & IoV crowd sensing (Generated) & Location & Low query failure; low comm. cost \\
\cite{ref107} & HE+\allowbreak{}BC~(D) & IoV crowdsourcing (Gowalla) & Location/\allowbreak{}Task & Multi-level privacy; high computation \\
\cmidrule{1-5}
\cite{ref108} & SMPC~(N) & Route planning (Simulation data) & Navigation & Optimal path selection; constant-time encryption \\
\cite{ref109} & SMPC+\allowbreak{}DP~(S) & Signal control (SUMO-generated) & Vehicle & Low impact on control performance \\
\cite{ref284} & SMPC+\allowbreak{}DP~(D) & Arrival-profile estimation (Shenzhen CV data + VISSIM) & Trajectory & Prevent inference attacks; 10.3 veh MAE/9.8\% MAPE; Low compute \\
\cite{ref285} & SMPC~(N) & Lane-change intention prediction (HighD) & Data & Two-server sharing; 92.84\% accuracy; MSE $\approx$1e-7 \\
\cite{ref286} & SMPC~(D) & Platooning coordination (Sim. Swedish highway) & Itinerary & Near fully-disclosed reward, no added overhead \\
\cite{ref110} & SMPC~(S) & Distributed data processing (Generated) & Data & Secure compression;99\%+ acc.; 94.96\% delivery \\
\cite{ref111} & SMPC+\allowbreak{}DP~(D) & Route privacy (TNTP) & Location/\allowbreak{}Route & Accurate travel-time estimation \\
\cmidrule{1-5}
\cite{ref112} & DP+\allowbreak{}FL~(C) & Traffic flow prediction (PeMS) & Data & Lowest attack success rate(2\%) \\
\cite{ref113} & DP~(S) & Traffic monitoring (Generated) & Communication & Balanced noise-computation trade-off \\
\cite{ref114} & DP~(C) & EV location privacy (Beijing charging posts) & Location/\allowbreak{}charging & 99\% query accuracy \\
\cite{ref115} & DP~(S) & Data streaming (TAPASCologne) & Data/\allowbreak{}Comm. & High utility on correlated data \\
\cite{ref274} & DP+\allowbreak{}FL~(C) & Trajectory prediction (NGSIM) & Parameter & Slow convergence(100 rounds = 77.55\% acc.) \\
\cite{ref275} & DP~(S) & Vehicle location publishing (SUMO/NS3) & Grid & Lower MAE; Fewer communication hops \\
\cite{ref276} & DP~(C) & Trajectory data publishing (X-Drive/K-Drive) & Trajectory & Lowest mutual-info. privacy leakage \\
\end{longtable}
}

\FloatBarrier

\subsubsection{Federated Learning}

FL is increasingly being utilized in intelligent transportation systems (ITS) to enhance privacy during downstream traffic management tasks such as mobility flow prediction (traffic or crowd), travel time estimation, traffic signal control and traffic status identification. Traditional ML models face challenges like data privacy risks, long training times, and high communication costs. FL mitigates these by enhancing privacy while improving computational and communication efficiency. FL for traffic flow prediction proposed in \cite{ref71} replaced the centralized road network model with a decentralized global GCN to balance accuracy and computational cost. Nonetheless, GCN-based models utilizing FL often neglect the underlying topological structure of the traffic networks, which can lead to potential privacy breach. FASTGNN \cite{ref72} is a potential solution to this which uses FL and GNN for traffic flow prediction that protects graph topological data privacy. The method also employs DP to protect privacy of local network structures as the global model is developed. Another development for traffic flow prediction is the ability to handle heterogeneous spatial characteristics across participants using vertical FL (VFL) \cite{ref73}. The authors in \cite{ref73} introduced a federated graph attention layer to preserve spatial information while capturing short-term temporal features. This allows participants with different spatial attributes to contribute to the global model training without privacy exposure.

Communication efficiency is a critical factor in FL-based traffic prediction systems. In \cite{ref74}, an improved FedAvg that uses random subsampling of participants has helped reduce communication costs by limiting the number of participants in each round. Additionally, clustering techniques based on geographical proximity or model similarity have been applied in \cite{ref75} to further minimize communication overhead. An asynchronous communication algorithm proposed in \cite{ref76} further enhances efficiency by adjusting the aggregation process to account for varying update times.

In addition to traffic prediction, FL is increasingly being utilized in various smart traffic management applications, such as crowd flow prediction, and traffic signal control. The authors in \cite{ref77} proposed an enhanced FL framework that integrates clustering algorithms to manage human trajectory data. Their method improved both the efficiency and accuracy of FL-based crowd mobility predictions. To address the challenge of predicting mobility across various participants with differing locations, a VFL framework for mobility prediction was introduced in \cite{ref78}. This framework facilitates joint learning over vertically partitioned data from multiple clients, without compromising sensitive data.

Traffic signal control is another key area where FL has been highly effective. Optimization of traffic signal control \cite{ref79} has utilized various ML techniques, however, these approaches do not account for hybrid traffic flow or vehicular control and coordination models. Reinforcement learning (RL) has also gained traction in this domain, but the high dimensionality of the joint action space still poses challenges for centralized RL in large-scale applications \cite{ref82}. Recent advancements have integrated FL and RL for traffic control systems where agents communicate remotely and load model parameters during off-peak hours, enhancing convergence speed of the algorithm \cite{ref83,ref84}. Federated deep RL has also been proposed in \cite{ref85} for network edge caching to improve QoS in VANETs.

Accurate vehicle positioning is essential for enhancing road safety and reducing traffic congestion. The authors in \cite{ref86} employed FL to analyze high-resolution remote sensing images to enhance traffic status identification and congestion monitoring systems while ensuring user privacy. Most remote sensing images or vehicle positioning data are retrieved from widely used GNSS and inertial navigation systems (INS). While these systems can provide accurate vehicle positioning, they are costly and heavily reliant on GNSS base stations. To overcome these challenges, authors in \cite{ref87} proposed the use of FL to enable cooperative positioning using fused information from both vehicles and infrastructure. This increased training samples by maximizing cooperation with all vehicles and provides high-precision positioning corrections without sharing location data.

Traditional travel mode identification and trajectory prediction depends on GPS-based segmentation and statistical features for classification \cite{ref88}. To reduce dependence on GPS data and provide user privacy, authors in \cite{ref90} proposed an FL-based system where smartphone inertial data is transferred to vehicles to infer the vehicle's position in real time without GPS signal. Another challenge for travel mode identification is the lack of labeled data, as users typically do not label their driving modes, and third-party applications do not collect such labels. To address this, \cite{ref91} proposed a semi-supervised FL framework that pseudo-labels local data using a small labeled cloud dataset. It aggregates data based on class distribution while keeping user data on device. Motivated by the scattering of pedestrian trajectory data across disconnected surveillance scenes, \cite{ref283} paired a destination-oriented long short-term memory (LSTM) encoder-decoder with a federated training scheme that lets multiple scene-specific clients collaborate without exposing local trajectories. Federated multi-scene training improves prediction accuracy over models confined to a single scene while preserving data privacy.

Addressing the heterogeneity of temporal patterns across distributed traffic clients, \cite{ref279} introduced a federated mixture-of-experts scheme that routes recurrent and irregular traffic dynamics to specialized local experts before synchronizing only the utilized experts across the network. Recognizing that federated traffic forecasting studies commonly assume a homogeneous sensing environment, \cite{ref281} identified spatial, temporal-coverage, and data-quality skew across regional clients and proposed a unified framework combining personalized clustering, time-window-based training, and joint global-local denoising. Since uniform aggregation breaks down under statistically heterogeneous urban traffic conditions, \cite{ref282} proposed an interpretable trust-aware federated forecasting framework that scores client reliability from validation performance and update behavior through a fuzzy inference system with temporal smoothing. The trust-calibrated aggregation accepts only a modest rise in forecasting error in exchange for more robust and transparent client weighting than standard FedAvg-based pooling.

\subsubsection{Blockchain-based PPML}

Various studies have integrated BC into ML for privacy-preserving ITS applications, including traffic management, traffic congestion control, traffic flow prediction, and destination prediction. For traffic management, authors in \cite{ref69} proposed a two-level privacy protection framework incorporating smart contract (SC)-based proof-of-work (PoW) to validate data integrity and counteract data poisoning threats and LSTM-autoencoder (LSTM-AE) to transform data into a secure, encoded format, mitigating inference attacks. Authors in \cite{ref93} proposed a BC-based model designed to estimate traffic congestion probabilities while preserving privacy. In \cite{ref94} a BC-enabled FL framework was proposed for urban traffic flow management, enhancing security by verifying model updates through miners to prevent malicious vehicles and reduce data poisoning. It also counteracts inference attacks using local DP (LDP) in the gradients. In another study \cite{ref95}, BFRT was introduced as a decentralized method for real-time traffic flow prediction. It uses permissioned BC technology to protect vehicle privacy while ensuring accurate predictions. A bi-level BC architecture was developed for secure FL-based traffic prediction in \cite{ref96}. It used a distributed HE federated averaging (DHFA) approach to secure the federated process and connect decentralized model validation.

Similar to traffic flow prediction, the authors in \cite{ref97} proposed a BC-integrated FL framework for destination prediction, addressing user and location privacy concerns. Besides specific IoV applications, BC-based FL can enhance the fairness and privacy of the ML frameworks. In \cite{ref98}, a decentralized architecture ensured privacy and fairness through peer-to-peer (P2P) collaboration. It employs a three-layer encryption combining DP, BC, and HE to protect data and improve accuracy, along with a credibility evaluation system to promote fairness and incentivize participants.

In collaborative ML applications, establishing trust among participants and ensuring the credibility of task publishers is crucial. The authors in \cite{ref99} proposed a hierarchical trust evaluation strategy for 5G-ITS, using a heterogeneous BC and ML. It assessed trustworthiness through three levels, with rewards or penalties, and reassessed trust after task completion. In \cite{ref100}, many-to-one matching model (based on reputation) was proposed to enable secure and efficient assignment of tasks. Authors in \cite{ref101}, proposed a two-tier BC architecture for FL to improve security and efficiency using local and global model update chains. Privacy preservation was obtained through task fragmentation, decentralized reputation management, and incentives, ensuring minimal data sharing and motivating honest participation.

To reconcile large-scale traffic-data aggregation with individual privacy, Chen et al. \cite{ref270} embedded ZKPs into a consortium BC for C-V2X, letting vehicles certify the validity of their telemetry without exposing the underlying values. Addressing the coexistence of internal-combustion and electric fleets, Dong et al. \cite{ref271} coupled a consortium-BC speed advisory system with Paillier homomorphic aggregation, allowing the base station to compute carbon-optimal recommendations directly on encrypted emission data submitted by vehicles. A tunable coordination factor further lets policymakers rebalance optimization weight between direct and indirect emissions as grid carbon intensity varies across regions and time. Wang et al. \cite{ref277} combined asynchronous federated aggregation with a BC-based dynamic-grouping PBFT consensus for multi-vehicle trajectory prediction, so that locally trained GRU-based models can be validated and fused across RSUs without exposing raw sensor histories. The resulting framework is shown to remain markedly more robust to gradient-tampering attacks than a non-BC federated counterpart. To secure hierarchical intelligence in IoV without relying on a trusted central server, \cite{ref278} designed a fully decentralized BC-FL framework in which base stations perform validation-based local selection while a two-tier microblock and keyblock structure coordinates global consensus. The scheme sustains strong recognition performance on the GTSRB benchmark even under adversarial label-flipping and partial base-station outages, addressing the single point of failure inherent to conventional BC-FL integrations.

Authors in \cite{ref269} targeted the tension between immutable accident records and the "right to be forgotten" by designing a decentralized attribute-based chameleon hash that lets multiple authorities jointly authorize block- and transaction-level redactions on a permissioned ledger. Since outsourcing raw crowdsourcing task metadata to a public ledger risks severe privacy exposure, Yang et al. \cite{ref273} devised a constant-length ciphertext attribute-based encryption scheme that keeps on-chain access-list storage proportional only to the number of task requesters rather than every authorization pair. A hybrid searchable-encryption matching protocol further supports keyword and range queries over encrypted tasks, closing a gap left by earlier BC crowdsourcing frameworks limited to single-keyword search. Given the semi-trusted nature of roadside units in vehicular edge networks, Li et al. \cite{ref272} paired FHE with a BC-recorded offloading ledger, and introduced a joint cost metric spanning energy, delay, and privacy exposure risk to steer task placement decisions. A self-learning potential-game algorithm then lets vehicles independently converge on dependable offloading strategies without relying on a central coordinator, outperforming heuristic and centralized RL baselines.

\subsubsection{Homomorphic Encryption}

HE has been applied for IoV applications such as traffic flow prediction, travel time estimation, and traffic signal optimization. Study \cite{ref102} used HE to secure model parameters, preventing inference attacks for traffic flow prediction. A prior study \cite{ref103} introduced a privacy-preserving aggregation scheme for vehicular fog computing using homomorphic threshold cryptosystems. While these studies analyzed security against honest-but-curious fog servers and dishonest users, they did not address advanced inference attacks or evaluate the reliability of secure model updates. In \cite{ref104}, BC, elliptic-curve cryptography (ECC), and FHE were used for secure, reliable model updates and privacy-preserving data transmission. Study \cite{ref96} used HE for privacy preservation and DHFA for secure aggregation of encrypted local updates. It also featured a partial private key distribution protocol and a partial HE/decryption scheme for robust privacy protection. To harden parameter aggregation against reverse inference and poisoning in heterogeneous connected-vehicle fleets, another study \cite{ref280} combined multi-key partially HE with secret sharing and a Gaussian noise mechanism so that the server can filter corrupted updates without accessing plaintext gradients.

To address the privacy and security concerns associated with real-time traffic and vehicle data from a large number of participants, \cite{ref105} proposed a fog-based vehicle crowd sensing (FBVC) architecture using HE. It features a two-tier fog framework with static upper-tier and dynamic lower-tier fog nodes, supporting secure data fusion, traceability, and integrity. Similarly, \cite{ref106} presented a privacy-preserving data sharing scheme for IoV, using modified Paillier cryptosystem to ensure location privacy and secure data aggregation. Proxy re-encryption enabled secure edge querying, providing strong privacy and resistance to collusion attacks. Additionally, \cite{ref107} introduced a decentralized location privacy-preserving scheme for spatial crowdsourcing using BC and HE. It also incorporated order-preserving encryption and non-interactive ZKP to prevent location forgery.

\subsubsection{Secure Multi-party Computing}

IoV applications such as traffic navigation, traffic signal control, and routing services with location privacy, leverage SMPC to protect sensitive vehicular data. The authors in \cite{ref108} proposed EPNS, a privacy-preserving IoV solution using SMPC and multiparty delegated computation (MPDC). It employs two non-colluding servers -- one for re-encryption and another for computation -- to prevent data leakage even if one is compromised. A traffic signal control framework for connected autonomous vehicles (CAV) was proposed in \cite{ref109}. The framework integrates SMPC and DP to protect CAV data against three types of privacy threats: collusion attacks, control center database attacks, and inference attacks. The framework used a linear programming model and arrival rate estimator to ensure efficiency in varying traffic scenarios. A two-stage stochastic programming model mitigates DP-induced noise impacts on performance.

For cycle-based arrival profile estimation drawn from multiple competing mobility companies, \cite{ref284} cast the problem as a matrix completion task in which additive secret sharing and Laplace perturbation jointly protect each company's connected-vehicle trajectories during co-reconstruction. To enable driver-assistance systems to anticipate lane changes without exposing surrounding vehicle telemetry, \cite{ref285} deployed a two-server SMPC protocol, including Beaver-triple multiplication and a secure tanh activation, around a recurrent neural network trained on the HighD dataset. Experiments show that encrypted inference reproduces plaintext accuracy almost exactly, though runtime remains bottlenecked by the least capable participating server. In the absence of a trusted broker for coordinating platoons across competing logistics carriers, \cite{ref286} developed a real-time coordination framework in which distributed additive secret sharing and gossip-based averaging let carriers jointly optimize hub waiting times without disclosing itinerary information to peers or third parties.

While SMPC-based cryptographic protocols provide privacy-preserving computation, leveraging distributed edge architectures can enhance scalability and computational efficiency. The SPEED framework in \cite{ref110} distributed data processing across edge nodes, reducing centralized attack risks. It also used compressed sensing, data shuffling, and split computation to minimize raw data exposure. Authors in \cite{ref111} proposed a privacy-preserving decentralized routing service using SMPC and DP. It combined additive and Shamir secret sharing to protect location data and the DP Laplace mechanism to enhance privacy in sparse data scenarios, reducing risks of tracing individual trajectories.

\subsubsection{Differential Privacy}

In ML-based IoV applications DP has been adopted in traffic flow prediction, real-time traffic monitoring, trajectory prediction and location privacy. Study \cite{ref112} combined FL and LDP for privacy-preserved traffic flow prediction. Their method applies Gaussian noise to the gradients of an LSTM model before sharing them with a central server for aggregation, preventing inference attacks. Furthermore, predicting and managing traffic flow effectively requires robust monitoring systems. Authors in \cite{ref113} presented a distributed traffic monitoring system that protects individual privacy using DP. They introduce three algorithms with different trade-offs between noise and computation, with the third approach providing the most balanced performance.

Besides traffic management, location privacy is crucial for protecting personal identification and routes. Authors in \cite{ref94} applied LDP to gradient updates in a BC-enabled FL framework for traffic flow prediction, protecting against location-revealing inference attacks. For EV location data in vehicle-to-grid (V2G) networks, a privacy-preserving mechanism using DP, quadtree spatial decomposition, and Bernoulli random sampling was proposed in \cite{ref114}. This approach effectively protected location privacy while minimizing relative errors in data utility. Study in \cite{ref115} developed a DP data streaming system for CAV networks. It addressed the challenges of data correlation and dynamic network topology by implementing group-based data compression and adaptive noise addition. Study \cite{ref274} confronted the residual leakage risk that persists in federated trajectory prediction despite avoiding raw-data sharing, by injecting Gaussian differential-privacy noise into locally trained parameters before they reach the aggregation server.

For communication with roadside units for traffic management, rather than assuming a fully trusted roadside unit, study \cite{ref275} proposed a location-publishing mechanism that limits how long any single vehicle can act as group leader, using directional thresholds and lifespan-based hopping to curb prolonged data collection under semi-trusted infrastructure. Study \cite{ref276} noted that generative trajectory synthesis alone can still closely mirror real user movement, and responded with a two-layer pipeline in which a density-aware LSTM-GAN first substitutes real trajectories with synthetic ones before a personalized differential-privacy step perturbs points according to local density and distance from cluster cores. Comparisons against related baselines showed the added perturbation layer meaningfully reduced mutual-information leakage without sacrificing trajectory data utility.

\subsection{Autonomous Driving and Safety-critical Applications}\label{sec:ad}

Autonomous driving and safety-critical applications require ultra-reliable and low-latency decision-making capabilities. These capabilities are powered by ML algorithms trained on vast datasets from in-vehicle communication and vehicle-to-everything (V2X) networks. PPML techniques are essential for ensuring privacy and mitigating adversarial attacks in autonomous driving systems, while also enabling efficient and reliable operation. We review the recent advances in the adoption of PPML techniques in this domain and the technical summary is given in Table~\ref{tbl:ad-summary}.

\subsubsection{Federated Learning}

Connected autonomous vehicles rely heavily on real-time data processing and decision-making such as monitoring system to ensure safety and efficiency. FL plays a pivotal role in supporting such tasks by enabling privacy-preserving collaboration across vehicles, infrastructure and environment. Studies have proposed in-vehicle driver monitoring systems that predict driver behavior, such as an upcoming lane change, to provide safety-critical alerts \cite{ref118}. However, these systems face significant computational and communication challenges. The FL approach proposed in \cite{ref121} utilizes the FedGKT to improve bandwidth efficiency through asynchronous training for driver activity recognition. Furthermore, driver-related data is often tied to personal habits, cultural nuances, and emotional states, making generalization across individuals inherently challenging. To address this, personalized FL technique in \cite{ref123} enhanced model adaptability to diverse driver behaviors, ensuring secure data handling while considering driver patterns and requirements. The authors in \cite{ref124} integrated HE scheme with FL to add an extra layer of privacy in detecting driver drowsiness. Additionally, advances like DFL framework and gossip protocols in \cite{ref126}, can improve model training in dynamic multi-user environments like passenger monitoring in public transit.

Another in-vehicle monitoring system is steering angle prediction, which is vital for self-driving and ADAS features like lane-keeping. The authors in \cite{ref129} used multi-modal data, including road images and optical flow, to enhance accuracy of steering angle prediction models. Another study \cite{ref130} optimized edge model quality by adding noise and incorporating various data sources across vehicles. The FL model also reduced communication overhead and achieved robustness to network disruptions. Targeting on-vehicle real-time perception, \cite{ref308} trained customized YOLOv8 detectors locally on each vehicle's onboard GPU and introduced a selective client-update mechanism that filters outlier contributions before FedAvg aggregation to cope with non-IID driving data. The framework maintains competitive detection accuracy while markedly lowering communication and computational costs.

FL is increasingly used in CAVs for collaborative training and improved motion controller parameters while ensuring data privacy. The proposed scheme in \cite{ref134} allowed CAVs to update controller parameters in real time, and continuously improved target speed achievement and vehicle handling through aggregated data. For collision avoidance and safe braking, FL has proven effective in optimizing collaborative control parameters among multiple CAVs at intersections without compromising privacy \cite{ref135}. Similarly, \cite{ref136} used FL for safe braking by improving the estimation of road friction coefficients in varied driving conditions and environments. Addressing the unfairness that arises when CAVs with heterogeneous data volumes and channel conditions perform identical local iteration counts, \cite{ref298} devised a personalized FL scheme that allocates local training rounds per vehicle to balance time and energy costs. To counter collusive adversaries that manipulate group-formation decisions among connected autonomous vehicles, \cite{ref305} built a federated trust-scoring mechanism that identifies malicious participants without requiring vehicles to expose raw sensing or trajectory data. In the context of vehicle-road cooperation, \cite{ref299} proposed SecureFL, which pairs an ensemble-based detector for federated gradient sign attacks with a graph neural network-driven reputation scheme that down-weights unreliable participants. 

Accurate vehicle trajectory prediction enables CAVs to plan movements, anticipate risky behaviors, and prevent accidents. FL framework can learn nuanced spatiotemporal features when combined with transformers \cite{ref139} and LSTM models \cite{ref140}, enhancing predictive accuracy while preserving privacy. Additionally, FL has been used for anomaly detection in vehicular trajectories. In \cite{ref141}, FL was integrated with the one-class support vector machine (OC-SVM) for anomalous driving behavior detection at intersection points. This approach supported continuous traffic adaptation, improving detection accuracy while preserving privacy. Furthermore, various studies \cite{ref142,ref143,ref144} showed that FL performs comparably to centralized learning while preserving privacy throughout the learning process. To further optimize trajectory prediction, authors in \cite{ref145} proposed personalized FL to enhance generalization in diverse driving environments. Meanwhile, \cite{ref146} employed a map fusion in three stage (density-based spatial clustering, score-based averaging, and intersection-over-union-based pruning) to enable accurate, privacy-preserving trajectory predictions.

{\scriptsize
\setlength{\tabcolsep}{3pt}
\begin{longtable}{@{}>{\raggedright\arraybackslash}p{0.055\linewidth}>{\raggedright\arraybackslash}p{0.095\linewidth}>{\raggedright\arraybackslash}p{0.33\linewidth}>{\raggedright\arraybackslash}p{0.13\linewidth}>{\raggedright\arraybackslash}p{0.33\linewidth}@{}}
\caption{Literature adopting PPML in autonomous driving and safety-critical applications.}\label{tbl:ad-summary}\\
\toprule
\textbf{Ref.} & \textbf{Tech. (Trust)} & \textbf{Application (Dataset)} & \textbf{Privacy target} & \textbf{Notable Outcome} \\
\midrule
\endfirsthead
\caption[]{Literature adopting PPML in autonomous driving and safety-critical applications. (continued)}\\
\toprule
\textbf{Ref.} & \textbf{Tech. (Trust)} & \textbf{Application (Dataset)} & \textbf{Privacy target} & \textbf{Notable Outcome} \\
\midrule
\endhead
\midrule
\multicolumn{5}{r}{\textit{Continued on next page}} \\
\endfoot
\bottomrule
\multicolumn{5}{@{}p{\linewidth}@{}}{\scriptsize Tech.: primary technique first, then any combined techniques (hybrid PPML). Trust assumption in brackets: C central aggregator, S semi-trusted edge/fog/RSU/cloud, N non-colluding servers, D decentralized (no trusted party).} \\
\endlastfoot
\cite{ref118} & FL~(C) & Lane-change prediction (Unity simulation) & Behavioral & 4\% higher accuracy; 7.6x training efficiency \\
\cite{ref121} & FL~(C) & Activity recognition (AI City Challenge/StateFarm) & Data/\allowbreak{}edge & 98.9\% recognition rate; Reduced resource \\
\cite{ref123} & FL~(C) & Driver monitoring (SFDDD/DriveAct) & Local/\allowbreak{}model & Large accuracy gain; 37.46\% lower comm. cost \\
\cite{ref124} & FL+\allowbreak{}HE~(C) & Drowsiness detection (NTHU-DDD/YAWDD) & Parameter & Improved comm. \& scalability \\
\cite{ref126} & FL~(D) & Driving recognition (AI City Challenge/StateFarm) & Data/\allowbreak{}driver & High accuracy; low comm./comp./storage \\
\cite{ref129} & FL~(C) & Steering angle prediction (SullyChen) & Data & 25\% less bandwidth, 75\% less training time \\
\cite{ref130} & FL~(C) & Steering angle prediction (CarND/Udacity sim) & Vehicle & Enhanced accuracy; 62--250x lighter network load \\
\cite{ref308} & FL~(C) & Object detection (KITTI/NuImages+Cityscapes) & Sensor & mAP50 84.3\%(IID)/61.6\%(non-IID); Low comp. cost \\
\cite{ref134} & FL~(C) & CAV motion control (BDD/DACT) & Data/\allowbreak{}Model & Superior velocity tracking; 40\% faster convergence \\
\cite{ref135} & FL~(C) & Motion control \& collision avoidance (N/A) & Data/\allowbreak{}vehicle & Motion accuracy; 0.44--12.80\% lower comm. cost \\
\cite{ref136} & FL~(C) & Safe braking (Generated) & Data & 2.58\% better braking acc.; 63\% less comm. overhead \\
\cite{ref298} & FL~(C) & CAV steering prediction (CARLA sim) & Data & Detects malicious CAVs in 3 rounds; 50\% lower time \\
\cite{ref305} & FL~(C) & AV group formation (NS3/SUMO sim) & Trajectory/\allowbreak{}sensing & 85.3\% malicious-node detection rate \\
\cite{ref299} & FL~(C) & AIoT vehicle-road cooperation security (WSN-DS) & Data & About 99\% classification accuracy \\
\cite{ref139} & FL~(C) & Vehicle trajectories (GAIA open dataset) & Spatial-temporal & High accuracy; Efficient communication \\
\cite{ref141} & FL~(C) & Trajectory anomaly detection (INTERACTION) & Data/\allowbreak{}trajectory & 98--99\% accuracy \\
\cite{ref142} & FL+\allowbreak{}HE~(C) & Trajectory prediction (NGSIM) & Model/\allowbreak{}parameter & Over 99\% accuracy; high overhead (FAHE) \\
\cite{ref143} & FL~(C) & Driving decision (Motion prediction dataset) & Data & Trust verification; 95\% accuracy; Real-time \\
\cite{ref144} & FL~(C) & Human trajectory prediction (ETH/UCY) & Model/\allowbreak{}design & Superior accuracy; better search efficiency \\
\cite{ref145} & FL~(C) & Trajectory prediction (CARLA synthetic) & Model/\allowbreak{}data & 2x better than baseline \\
\cite{ref146} & FL~(C) & Contextual trajectory prediction (CARLA sim) & Map & High fidelity; real-time fusion \\
\cite{ref153} & FL~(C) & Object detection (KITTI) & Data & Comparable to centralized (68\% mAP) \\
\cite{ref154} & FL~(C) & Snow detection (CADC) & Data/\allowbreak{}model & Enhanced detection; real-time \\
\cite{ref155} & FL~(D) & Road user classification (nuScenes) & User & Outperforms self-learning methods \\
\cite{ref156} & FL~(C) & 3D pothole detection (Generated) & Crowd/\allowbreak{}spatial & High detection accuracy; lightweight \\
\cite{ref157} & FL~(C) & Road damage detection (Generated) & Data/\allowbreak{}user & Warning accuracy; 75\% lower comp./comm. cost \\
\cite{ref158} & FL~(C) & Surface classification (RSCD) & Data & Classification security; low comm. cost \\
\cite{ref159} & FL~(C) & Road damage (JRDD/MRDD) & Regional/\allowbreak{}local & 1.33--163\% accuracy improvement \\
\cite{ref160} & FL~(C) & License plate recognition (Fujian traffic police) & Data/\allowbreak{}network & High accuracy; low latency \\
\cite{ref161} & FL~(C) & Traffic sign recognition (BelgiumTS) & Model & Superior accuracy; high energy efficiency \\
\cite{ref162} & FL~(C) & Resource optimization/perception (CarlaFLCAV) & Sensor/\allowbreak{}model & High perception accuracy; low network resources \\
\cite{ref164} & FL~(C) & Resource optimization (MNIST/BelgiumTSC) & Model/\allowbreak{}vehicle & Better aggregation accuracy \\
\cite{ref165} & FL~(C) & Resource optimization (FLIR/BDD100K) & Training & 60\% less training time, 80\% less bandwidth \\
\cite{ref301} & FL~(D) & Privacy-preserving FL framework (MNIST/CIFAR-10) & Model & 98.97\% accuracy; 90\% acc. in 4 rounds \\
\cite{ref302} & FL~(C) & Bird's-eye-view perception (CARLA/OpenCDA sim) & Position/\allowbreak{}model & Over 50\% IoU improvement \\
\cite{ref304} & FL~(C) & 3D object detection (V2X-Fed sim) & Data/\allowbreak{}parameter & Retains 97.51\% of centralized accuracy \\
\cite{ref306} & FL~(C) & Vehicle/object counting (CARPK et al.) & Data & MAE 6.08/RMSE 8.42 (best compared); Lightweight \\
\cmidrule{1-5}
\cite{ref166} & BC~(D) & Route optimization \& emergency response (Generated) & Data & Reliable model validation; Efficient collaboration \\
\cite{ref167} & BC~(D) & Autonomous decision making (Generated) & Decision & High decision accuracy; Adaptive learning \\
\cite{ref168} & BC~(D) & Vehicle positioning (Generated) & Location & Lowest positioning error; Real-time \\
\cite{ref291} & BC~(D) & Real-time vehicle intrusion monitoring (Ethereum testbed sim) & Vehicle data & Encrypted registration keys; 94\% detection accuracy; about 3000 req/min \\
\cite{ref169} & BC+\allowbreak{}FL~(D) & Misbehavior detection (Generated) & Model security & 97\% accuracy; Low latency/energy consumption \\
\cite{ref170} & BC+\allowbreak{}HE~(D) & Autonomous driving (Rancho Palos Verdes/San Pedro) & Identity/\allowbreak{}message & 99\% privacy; 73.7\% lower training loss \\
\cite{ref171} & BC+\allowbreak{}DP~(D) & Misbehavior detection (VeReMi) & Data & High detection accuracy; Resource efficiency \\
\cite{ref172} & BC+\allowbreak{}FL~(D) & Privacy leakage prevention (IoT-Botnet/ToN-IoT) & Data & High detection rate; Resource efficiency \\
\cite{ref173} & BC+\allowbreak{}FL~(D) & Intrusion detection (KDDCup99) & Storage/\allowbreak{}sharing & High detection rate; low resource use \\
\cite{ref174} & BC+\allowbreak{}FL~(D) & Intrusion detection (Car-Hacking/ToN-IoT) & BC update & High attack detection; Reliability \\
\cite{ref292} & BC+\allowbreak{}FL~(D) & Intrusion detection (CAR-Hacking/UNSW-NB15) & Data & Collective defense; 99.32\%/99.69\% accuracy \\
\cite{ref314} & BC+\allowbreak{}FL~(D) & Misbehavior / intrusion detection (VeReMi) & Model & Secure aggregation; 68--94\% precision, 57--88\% recall \\
\cite{ref287} & BC+\allowbreak{}FL~(D) & Hierarchical aggregation (CIFAR-10/FashionMNIST) & Model & Off-chain training; 65.54\% faster to target accuracy \\
\cite{ref289} & BC+\allowbreak{}FL~(D) & Driving distraction recognition (State Farm) & Data & Prevent leakage; 93\% FL acc.; 4736 tx/s throughput \\
\cite{ref290} & BC+\allowbreak{}HE~(D) & Malicious vehicle detection (SUMO VANET sim) & Reputation ratings & Encrypted reputation; +38.56\% detection rate \\
\cite{ref288} & BC+\allowbreak{}FL+\allowbreak{}DP~(D) & Privacy-preserving data sharing (MNIST sim) & Data & Resist inference attack; 99.3\% accuracy; $\sim$0.09s consensus delay \\
\cmidrule{1-5}
\cite{ref175} & HE~(C) & Distracted-driver detection (State Farm) & Data/\allowbreak{}service & 86.2\% accuracy; long classification time \\
\cite{ref177} & HE+\allowbreak{}FL~(C) & Driver fatigue detection (UTA-RLDD) & Driver & 60--96\% less computation, 95\% less traffic \\
\cite{ref300} & HE+\allowbreak{}FL~(S) & General-purpose FL training (VeReMi) & Model updates & Verified under Dolev-Yao model; 59\% less computation \\
\cite{ref311} & HE+\allowbreak{}FL~(S) & Secure aggregation (MNIST/FMNIST/CIFAR-10) & Gradients & Resist model-recovery attack; under 1KB overhead \\
\cite{ref312} & HE+\allowbreak{}FL~(C) & Collusion-resistant verifiable FL (MNIST/CIFAR-10) & Gradients & Resist collusion attacks; Low runtime \\
\cite{ref303} & HE+\allowbreak{}FL~(C) & FL against poisoning (Car-Hacking + 4 datasets) & Model & Last-layer encryption; Up to 94.3\% accuracy \\
\cite{ref309} & HE+\allowbreak{}FL+\allowbreak{}BC~(D) & Misbehavior / anomaly detection (CICIoV2024/5G-NIDD) & Model updates & End-to-end encrypted updates; 99.9\% accuracy \\
\cite{ref310} & HE+\allowbreak{}FL~(D) & Cross-trust-domain collaborative learning (CIFAR-10) & Model & Resist gradient inversion; 7.6x faster encryption \\
\cmidrule{1-5}
\cite{ref179} & SMPC~(D) & CAV control (NGSIM) & State/\allowbreak{}input & Safe/optimal control; under 30ms overhead \\
\cite{ref180} & SMPC+\allowbreak{}FL~(C) & Vehicle classification/attack prevention (BIT-Vehicle) & Terminal/\allowbreak{}model & High prediction accuracy \\
\cite{ref181} & SMPC~(D) & Speed advisory (SUMO sim) & Data & Optimal advisory; single-iteration convergence \\
\cite{ref307} & SMPC+\allowbreak{}HE~(N) & Vehicle misbehavior/intrusion detection (MNIST/NSL-KDD) & Model/\allowbreak{}identity & Over 98\% accuracy; 16x faster upload \\
\cite{ref313} & SMPC~(N) & Cross-manufacturer training (benchmark networks) & Proprietary data & 3-party computation; Up to 907x faster inference \\
\cite{ref315} & SMPC~(N) & Privacy-preserving inference (benchmark DL models) & Data/\allowbreak{}model & Low comp./comm. overhead; 86x faster \\
\cite{ref316} & SMPC~(N) & Privacy-preserving traffic signal control (Chania sim) & Occupancy data & Encrypted values; about 0.13s overhead/control-cycle \\
\cmidrule{1-5}
\cite{ref182} & DP~(C) & In-vehicle monitoring \& driving statistics (BROOK) & Data/\allowbreak{}route & Improved utility; 95.69\% less info. loss \\
\cite{ref297} & DP+\allowbreak{}FL~(C) & In-vehicle intrusion (CAN-Intrusion + 2 datasets) & Model & F1 0.975--0.985; Reduced comm. cost (54MB) \\
\cite{ref183} & DP~(C) & Trajectory publishing (Montreal transit) & Trajectory & Improved runtime and scalability \\
\cite{ref184} & DP~(C) & Enhanced trajectory partitioning (SUMO sim) & Community & Reduced info. loss; Outperforms k-anonymity \\
\cite{ref185} & DP~(C) & Vehicle trajectory (GeoLife/Shanghai) & Location & density-weighted budget allocation \\
\cite{ref187} & DP~(D) & Intrusion detection (NSL-KDD) & Training & Multi-level security; 96.94\% accuracy \\
\cite{ref188} & DP+\allowbreak{}FL~(C) & Intrusion detection (VeReMi Extension) & Model & 0.95--0.97 F1-score; Distributed processing \\
\cite{ref293} & DP+\allowbreak{}FL~(S) & Intrusion detection (CICIDS-2017/CarHacking20) & Client/\allowbreak{}RSU & Lowest per-round ($\sim$36~MB/round) wait \\
\cite{ref295} & DP+\allowbreak{}FL~(C) & CAN intrusion detection (Car-Hacking + 2 datasets) & Data & 97.6\% accuracy; Reduced 40\% communication \\
\cite{ref294} & DP+\allowbreak{}FL~(D) & Traffic sign/object detection (GTSRB + 4 datasets) & Exchange & Acc. degrades as vehicles scale \\
\cite{ref296} & DP+\allowbreak{}FL~(C) & Autonomous EV model training (Udacity) & Model & Resist membership attack; Highest accuracy \\
\cite{ref189} & DP+\allowbreak{}BC~(D) & Secure communication (N/A) & Vehicle & System reliability; Resource efficiency \\
\end{longtable}
}

Computer vision for accurate vehicle, pedestrian and obstacle detection is crucial for safety and autonomous operations. FL enhances computer vision tasks by enabling decentralized data processing, preserving privacy, and reducing communication overhead while improving detection through collaborative learning. The study in \cite{ref153} demonstrated that FL architectures enhance object recognition at image boundaries by sharing aggregated model weights across vehicles. For instance, collaboratively trained YOLO improved detection of distant objects like trucks and pedestrians using indirect data from other users. Authors in \cite{ref154} proposed YOLO-CNN to improve CAV safety in snowy conditions. Additionally, study \cite{ref155} proposed a decentralized FL framework for object classification in CAVs using LiDAR data. PointNet model parameters are exchanged over V2V networks, reducing data centralization and preserving privacy.

Besides objects, pedestrians and climate, poor road conditions can pose serious risks to autonomous driving, traffic safety, and vehicle integrity. Authors in \cite{ref156} used FL for pothole detection, combining 3D-FL and YOLO for accurate defect size estimation. This model could also accurately distinguish real potholes from patched areas and artificial road bumps, countering adversarial ML attacks while preserving privacy. Similarly, \cite{ref157} proposed an adaptive FL framework for detecting hazardous road damage, classifying it by severity. Additionally, \cite{ref158,ref159}, explored various road surface condition and damage detection using distributed FL.

Object detection applications, such as license plate detection/recognition (LPR) and traffic sign recognition provide high quality dataset for the downstream ML tasks. LPR involves detecting the plate and recognizing its characters, however, challenges such as multi-directional detection and motion blur remain in these applications. The authors in \cite{ref160} redesigned the LPR model for edge devices using privacy-preserving FL, addressing issues such as blur and orientation discrepancies and introduces a tilt correction algorithm to enhance model robustness. For traffic sign recognition, study in \cite{ref161} proposed FL with a spiking neural network based on receptive fields that significantly reduced computational overhead and also mitigates the risk of location-based privacy leaks inherent in these systems. For collaborative bird's eye view perception among vehicles with heterogeneous camera configurations, \cite{ref302} kept the transformer's camera-dependent positional embeddings private to each client while sharing the remaining encoder, attention, and decoder parameters for aggregation.

To reduce computation overhead and improve FL performance for object detection and other applications, studies have proposed various techniques such as multistage resource allocation and strategic vehicle selection to optimize resource use and network bandwidth \cite{ref162}. Selective model aggregation proposed in \cite{ref164}, also improved performance by integrating high-quality models from well-resourced clients. To further optimize model selection authors in \cite{ref165} introduce an asynchronous federated aggregation protocol specifically for target recognition. Similar to \cite{ref164} this protocol also selects optimal local models based on data quality and computational capacity of clients. In another study \cite{ref304}, rather than exchanging full backbone weights, the study equipped a shared 3-D detection network with lightweight adapter modules so that cooperating vehicles only fine-tune and transmit a small parameter subset during federated training. The framework preserves detection performance close to full centralized fine-tuning while sharply reducing the communication payload exchanged among vehicles. For resource-constrained roadside and in-vehicle counting devices, \cite{ref306} coupled a lightweight attention-enhanced convolutional architecture with a model-contrastive federated objective that stabilizes training across non-IID client distributions without centralizing image data. Rather than relying on a trusted central server, \cite{ref301} proposed FLAV, which routes model updates through multiple parallel chains of vehicles carrying randomly masked parameters and filters out contributions that deviate excessively from the aggregated chain history.

\subsubsection{Blockchain-based PPML}

BC-PPML in CAV applications ensures secure and decentralized data sharing and computation. Authors in \cite{ref166} introduced a collective learning architecture where each CAV can independently train local ML models for tasks like route optimization and emergency response. The architecture allows CAVs to download validated models from the BC and facilitate efficient collaboration without compromising privacy or system reliability. The authors further combine ML with expert systems and deep RL in \cite{ref167} to enhance decision-making capabilities in autonomous driving scenarios. Study \cite{ref168} proposed a BC-integrated ML framework for vehicle positioning in CAVs. It uses a deep neural network (DNN) to predict GPS errors and store corrections on a BC to improve security and accuracy while preserving location privacy. For securing BC networks with real-time monitoring, \cite{ref291} integrated a bidirectional BC data pipeline with a signature-, anomaly-, and behavior-based intrusion detection that authenticates high-security registration plates and encrypts vehicle data before storage.

Study \cite{ref169} proposes a BC-supported FL framework to detect and mitigate malicious behaviors by distributing ML models across devices and secure training with BC consensus. Besides mitigating malicious vehicle behaviours, BC integration also ensured transparency and trustworthiness. Study \cite{ref170} leverages BC to incentivize honest participation and balance vehicular privacy with accountability using traceable identity-based schemes. The dual encryption mechanism ensured anonymous authentication for semi-honest vehicles and used BC-based traceability and reputation incentives to deter malicious behavior. Similarly, \cite{ref171} introduced a BC-based FL framework to detect and manage malicious participants. To address the privacy leakage, authors in \cite{ref172} proposed a dual-module system combining BC for secure transmission and DL for detecting malicious participants.

For intrusion detection, \cite{ref173} proposes a collaborative learning framework integrating FL and BC to enhance security, reduce latency, and protect privacy in vehicular edge computing. Expanding on this, \cite{ref174} proposed a DL-based intrusion detection system using vehicle nodes for distributed training. It consists of a converter network for attack classification and BC for reliable distributed training.For collaborative cyberattack detection across connected vehicles, \cite{ref292} proposed a BC-based FL scheme that selects the top-performing local models and counters class imbalance in intrusion traffic. Authors in \cite{ref314} presented a VANET misbehavior detection framework that trains several neural architectures under federated averaging and anchors the global model to a BC via smart contracts and hash fingerprints to guarantee its integrity and traceability. Across sixteen attack types, the collaborative model consistently surpassed its counterpart.
 
To curb the single point of failure and staleness that plague hierarchical FL, \cite{ref287} grouped clients by data-distribution similarity and applied synchronous aggregation within groups alongside asynchronous aggregation across them, with a lightweight BC layer restricting on-chain storage to hashes and verification results rather than full model parameters. The resulting scheme converges markedly faster than FedAvg-style baselines while modestly raising global accuracy. Targeting reputation-based trust management for FL in IoV, \cite{ref289} built a hierarchical BC architecture in which vehicle reputations are computed from direct and recommended trust and used to weight gradient aggregation, while a multi-sidechain proof of queue (PoQ) consensus mechanism replaces costly proof-of-work validation with accuracy-based verification. Rather than relying on a centralized reputation server, \cite{ref290} distributed vehicle trust computation across a BC of mobile edge servers and roadside units, encrypting reputation ratings with a multi-key FHE scheme so that collaborative scoring never exposes raw feedback. For secure vehicle-to-RSU data sharing, \cite{ref288} coupled FL with a dual DP mechanism and a lightweight local BC, injecting client-side noise across non-orthogonal data subsets before a dynamic gradient-aggregation step filters out poisoned updates.

\subsubsection{Homomorphic Encryption}

HE integrated into in-vehicle monitoring and secure communication systems, enable remote data processing while preserving user and vehicle privacy. Authors in \cite{ref175} utilized HE for secure neural network classification on encrypted data to detect distracted drivers. It employed FHE schemes and transciphering for efficient encryption, even in resource-constrained automotive settings. HE-friendly activation functions enabled privacy-preserving inference. To secure sensitive driver data during training, the authors in \cite{ref124} proposed a privacy-preserving framework for driver drowsiness detection (PFTL-DDD) utilizing FL. This framework employs Cheon–Kim–Kim–Song (CKKS) HE to protect data throughout the training process while transferring knowledge from a pre-trained model to a FL framework. Similarly, study \cite{ref177} introduced PRIV-DRIVE, which combines FL with Paillier HE (PHE) for driver fatigue detection. It encrypted model parameters using PHE and top-k selection, preserving data privacy while optimizing computation and communication.

To relieve the communication and single-point-of-failure burdens of conventional aggregation in the IoV, \cite{ref300} compressed client updates through low-rank approximation and quantization before encrypting them under a Paillier-based hierarchical scheme in which fog nodes aggregate ciphertexts without accessing plaintext. Study \cite{ref311} introduced a lightweight secure aggregation and verification scheme for cloud-edge collaborative IoV. It paired Chinese Remainder Theorem based single-masking protocol with a homomorphic hash verification mechanism that resists gradient-recovery attacks while keeping per-user overhead below comparable schemes. Another study \cite{ref312} proposed a verifiable secure aggregation protocol built on a client-initiator-server architecture that layers homomorphic pseudorandom masking and linear homomorphic hashing with Sigma-protocol proofs. Collusion tests show reconstructed gradients remain statistically indistinguishable from noise even under high proportions of colluding participants.

Addressing the coupling of poisoning and gradient-inference threats in IoV FL, \cite{ref303} confined HE to the last network layer and screened client contributions with an adaptive cosine-similarity threshold, keeping cryptographic overhead low while still filtering malicious updates. The scheme sustains higher accuracy than several robust-aggregation baselines under both label-flipping and inner-product-manipulation attacks. To reconcile end-to-end confidentiality with the real-time demands of vehicular anomaly detection, \cite{ref309} combined an adaptive CKKS scheme with dynamic precision scaling to curb ciphertext noise growth. The framework converged to near-perfect detection accuracy within a handful of rounds while keeping homomorphic overhead manageable through precision scaling and off-chain storage. Rather than relying on a central aggregator, \cite{ref310} organized vehicles into GAT-clustered ring topologies and protected locally trained models with an obfuscated modular-decomposition encryption scheme. Each ring member recovered the aggregated model through the Chinese Remainder Theorem without ever exposing plaintext parameters.

\subsubsection{Secure Multi-party Computing}

In the literature SMPC is used in applications such as cooperative control, cruise control, vehicle classification, and speed advisory systems. Study in \cite{ref179} proposed a privacy-preserving control framework for CAVs in mixed traffic using affine masking. Built on data-enabled predictive leading cruise control, it optimized traffic flow and improved fuel efficiency while ensuring rigorous privacy guarantees. Another study \cite{ref180} proposed FedVPS, a framework leveraging SMPC to enhance privacy and security in FL for IoV. By integrating SMPC with DP, FedVPS addressed non-IID data and model heterogeneity and used prototype-based aggregation to reduce communication costs. It also addresses privacy threats, including reconstruction and membership inference attacks. A privacy-preserving consensus-based speed advisory system (CSAS) using SMPC was introduced in \cite{ref181}. It enabled vehicles to compute an optimal consensus speed without exposing private data by securely sharing secret mappings, such as speed-emission data.

Recognizing that roadside units themselves can be compromised or dropped, \cite{ref307} distributed each vehicle's local update across multiple neighboring RSUs using packed secret sharing and let two non-colluding servers obliviously reconstruct and aggregate the encrypted shares with Paillier HE. Authors in \cite{ref313} developed a three-party secure computation framework that lets automobile manufacturers jointly train and run convolutional perception models for autonomous driving without exposing proprietary data. Another study \cite{ref315} presented a tripartite secret-sharing inference framework for autonomous driving that pairs a simplified private comparison and multiplication protocol with an adaptive fixed-point quantization scheme and a GPU library for integer matrix operations. Study \cite{ref316} proposed an SMPC-based traffic-responsive signal control strategy that secret-shares link-occupancy measurements across several semi-honest cloud servers and computes green-time plans directly on the encrypted shares. The method also detects and recovers from false-data-injection attacks on signal timings, reducing an attacker's ability to induce worst-case congestion.

\subsubsection{Differential Privacy}

In autonomous driving, in-vehicle monitoring and driving statistics collect a significant amount of sensitive information. Study in \cite{ref182} evaluates the application of DP for in-vehicle monitoring and driving statistics, balancing privacy and service quality. The authors proposed high utility (HUT) for batched queries which enhanced data utility while preserving privacy guarantees. Another study \cite{ref297} exploited the temporal and spatial homogeneity of driver behavior to structure CAN and network traffic into time-varying state matrices for a bidirectional temporal detector trained under FL. Laplace noise injected into the uploaded gradients prevents model inversion from recovering sensitive driving behavior while a random client-scheduling strategy stabilizes convergence. 

Real-time reporting in IoV systems raises privacy concerns, as vehicular trajectory data can reveal sensitive behavioral patterns. SafePath \cite{ref183} mitigated this risk using DP by constructing a noisy prefix tree for secure trajectory publication. Another study, \cite{ref184} integrated DP with exponential mechanism with trajectory partitioning and clustering to enhance efficiency and data utility while reducing information loss. Similarly, \cite{ref185} combines a dynamic sampling strategy with a Kalman filter, adding Laplace noise to balance data availability and privacy of vehicle trajectory.

Various CAV and traffic management systems use intrusion detection systems (IDS) to enhance security. However, data breaches can compromise IDS training to avoid detection. To address this, \cite{ref186} proposed a DP-based ML IDS for VANETs, using alternate-directional multipliers and dual variable perturbation to balance security and privacy. Similarly, \cite{ref187} introduced a secure and private IDS, which integrated DP with the alternating direction method of multipliers to protect V2V communication. Expanding on this, \cite{ref188} developed a DP-FL framework to prevent membership inference attacks while optimizing the utility-privacy trade-off. Addressing straggler-prone edge learning in vehicle-road collaboration systems, \cite{ref293} combined a CNN-Transformer IDS with an asynchronous FL protocol whose differential-privacy noise magnitude varies dynamically and the aggregation weights adapt to update staleness. To defend CAN bus intrusion detection against gradient inversion attacks in the IoV, \cite{ref295} designed a self-adaptive DP mechanism that dynamically tunes the per-round privacy budget according to real-time vehicle threat levels. To support decentralized model training without a central aggregator, \cite{ref294} developed a push-sum-based decentralized FL algorithm that perturbs local updates with DP and estimates gradients through a two-point method when explicit gradient access is unavailable. For autonomous electric vehicles, \cite{ref296} introduced an LDP-FL framework that perturbs local model updates with Gaussian noise before server-side aggregation. The scheme also resists malicious participants and inference-based attacks.

A few studies have integrated BC with DP to enhance privacy and trust of IoV systems. Authors in \cite{ref171} proposed a BC-based FL scheme with DP for misbehavior detection in VANETs. It leveraged DP with the Gaussian mechanism to provide strict privacy protection. Another study \cite{ref189} integrated privacy guarantees of LDP and distributed ledger technology of IOTA to achieve scalability, immutability, and quantum resistance in large-scale vehicular networks.

\subsection{Communication Infrastructure and Smart Services}\label{sec:infra}

Modern IoV ecosystems rely on robust communication infrastructures for smart services like EV charging, predictive maintenance, smart parking and context-aware infotainment. PPML techniques ensure data privacy while maintaining performance across heterogeneous networks. We review the recent advances of PPML techniques in this domain and the technical summary of relevant literature is given in Table~\ref{tbl:infra-summary}.

\subsubsection{Federated Learning}

Vehicular cyber-physical systems (VCPS) use V2X communication to integrate vehicles, infrastructure, and traffic management. These systems contain sensitive vehicular and user data. Authors in \cite{ref190} designed a FL model to prevent data leakage within VCPS. For anomaly detection within these systems, the study in \cite{ref191} proposed OES-Fed, which identifies and mitigates abnormal data inputs without transferring vehicle-specific data. Another adaptation is the application of extreme value theory (EVT) and personalized FL in \cite{ref192}, which address heterogeneous data distributions among vehicles by modeling rare, anomalous events while preserving privacy. The study in \cite{ref193} integrates deep Q-network (DQN) with FL to significantly reduce latency in vehicular data sharing. Furthermore, the resilience of FL to adversarial attacks in VCPS is enhanced through integration with other PPML techniques such as DP \cite{ref194} and BC-leveraged methods \cite{ref195}.

The smart parking control system is essential for managing urban parking but relies on third-party cloud storage. Authors in \cite{ref196} proposed a FL framework for real-time parking predictions, allowing vehicles to forecast availability without sharing sensitive data. With the rise of EVs, parking areas now integrate CSs managed through centralized systems to optimize charging, which can lead to system overload and privacy breaches \cite{ref47}. To avoid system overload, studies have integrated FL with BC and ML techniques for power load prediction \cite{ref197}, while clustering-based approaches have reduced communication costs and prediction bias \cite{ref198}. However, the diverse characteristics of EV and CS remain a challenge for collaborative learning. A cross-platform FL framework in \cite{ref199} combined recommendation models with encryption techniques to balance privacy and real-time prediction accuracy. Economic-driven FL model proposed in \cite{ref200}, further optimize multi-agent charging scenarios, where a multi-principal one-agent (MPOA) model transforms CS utility problem into a decentralized non-cooperative energy optimization to ensure privacy-aware resource sharing. Furthermore, to address unpredictable consumption patterns, study \cite{ref201} enhanced FedAvg with a probabilistic algorithm for better adaptability, while operators and aggregators balance cost, efficiency, and resource.

Route planning is an intelligent service which relies on centralized systems to gather information of complex road networks. To address the latency issues in these systems, \cite{ref202} proposed a FL-based decentralized approach using fog nodes and RSUs, reducing memory usage, latency, and communication overhead. Clustering technique in \cite{ref203}, balanced computational loads across edge nodes, enhancing data processing efficiency. Using A* algorithm on time-dependent graphs enabled accurate route selection while preserving privacy by minimizing cloud data transmission.

In addition to route planning, FL is increasingly applied in smart user applications like travel time estimation and destination prediction. For travel time estimation in \cite{ref204}, a global model is trained using data from all participants, while personalized models are fine-tuned for individual driving patterns. FL has also been used for cross-area travel time estimation, where localized models are trained in different regions and combined through FL to preserve privacy across geographic boundaries \cite{ref205}. Additionally, FL has been applied to destination prediction tasks, providing precise location services without exposing sensitive user data \cite{ref97}.

The expansion of IoV services within CAVs poses challenges in communication efficiency, energy use, and privacy. FL addresses these issues by selecting clients and servers efficiently during training \cite{ref209}, which significantly improved resource management and system responsiveness \cite{ref210,ref211}. Authors in \cite{ref212} proposed a CNN-based FL framework to enhance model quality through hierarchical aggregation at edge and cloud. The approach considered factors such as RSU proximity and vehicle density. In \cite{ref86}, FL overcame the limitations of single-node data processing without compromising the privacy of sensitive geospatial information. Furthermore, techniques such as EVT and Lyapunov optimization were employed to optimize FL frameworks, enabling better handling of anomalous events and dynamic power allocation \cite{ref213}.

Authors in \cite{ref327} moved away from server-based coordination by proposing decentralized synchronous and asynchronous FL variants for time series forecasting across EV charging ecosystems. The approach mitigated the single point of failure and straggler while lowering communication overhead. Another study \cite{ref337} tackled the single-point-of-failure risk of fixed aggregators by rotating the server role among vehicles according to a weighted score of training activity, credibility, speed stability, and distance. The rotating-server design suppresses selection of malicious clients and yields more stable convergence. Study \cite{ref336} combined FL with an Improved Football Team Training Algorithm to coordinate energy trading decisions among EVs and mobile CS under a time-of-use pricing scheme. The framework argued that lightweight public key infrastructure (PKI)-based encryption and locally trained models are sufficient to protect sensitive trading information in resource-constrained EV deployments. Study \cite{ref342} extended EV charging occupancy forecasting to more realistic setting of non-IID data collected independently across CS operators. The study found that variance-reduction and personalization-based aggregators cope substantially better with skewed feature distributions and unbalanced station utilization than plain averaging. Authors in \cite{ref343} introduced an aggregation rule that scales each client's momentum term according to its data volume, collection density, and feature richness to counter statistical heterogeneity across a fleet's battery telemetry.

Authors in \cite{ref334} proposed a mobility- and resource-aware cluster head election scheme for vehicular FL that incorporates a trust evaluation mechanism to discourage vehicles from evading aggregation duties. Study \cite{ref339} coupled over-the-air computation with an adaptive local-iteration mechanism so that vehicles with limited or fluctuating computing resources can still contribute meaningfully to each FL round rather than dropping out. Study \cite{ref340} proposed pairing top-k gradient sparsification with a Euclidean-similarity-based data-quality assessment and a reverse-auction-driven dynamic reputation mechanism to recruit high-quality, intermittently connected vehicles under limited communication budgets. To address straggler effects caused by vehicle mobility and heterogeneous connection quality, \cite{ref345} devised a semi-asynchronous federated scheme that clusters connected vehicles into network layers by expected completion time and weights aggregation by an age-of-information factor. Recognizing that vehicles are often reluctant to contribute resources to federated training, \cite{ref346} designed an online auction framework in which roadside infrastructure dynamically selects clients and sets rewards according to bidding behavior, delay tolerance, and model accuracy. The algorithm raised platform utility while lowering the cost to sustain client participation.

{\scriptsize
\setlength{\tabcolsep}{3pt}
\begin{longtable}{@{}>{\raggedright\arraybackslash}p{0.055\linewidth}>{\raggedright\arraybackslash}p{0.095\linewidth}>{\raggedright\arraybackslash}p{0.33\linewidth}>{\raggedright\arraybackslash}p{0.13\linewidth}>{\raggedright\arraybackslash}p{0.33\linewidth}@{}}
\caption{Literature adopting PPML in communication infrastructure and smart services.}\label{tbl:infra-summary}\\
\toprule
\textbf{Ref.} & \textbf{Tech. (Trust)} & \textbf{Application (Dataset)} & \textbf{Privacy target} & \textbf{Notable Outcome} \\
\midrule
\endfirsthead
\caption[]{Literature adopting PPML in communication infrastructure and smart services. (continued)}\\
\toprule
\textbf{Ref.} & \textbf{Tech. (Trust)} & \textbf{Application (Dataset)} & \textbf{Privacy target} & \textbf{Notable Outcome} \\
\midrule
\endhead
\midrule
\multicolumn{5}{r}{\textit{Continued on next page}} \\
\endfoot
\bottomrule
\multicolumn{5}{@{}p{\linewidth}@{}}{\scriptsize Tech.: primary technique first, then any combined techniques (hybrid PPML). Trust assumption in brackets: C central aggregator, S semi-trusted edge/fog/RSU/cloud, N non-colluding servers, D decentralized (no trusted party).} \\
\endlastfoot
\cite{ref190} & FL~(C) & Data leakage prevention in VCPS (20 Newsgroups) & Data & High accuracy; leakage detection \\
\cite{ref191} & FL~(C) & Outlier detection (MNIST/CIFAR-10/BAIDU) & Data/\allowbreak{}Model & Enhanced accuracy; reduced noise \\
\cite{ref192} & FL~(C) & Anomalous event detection (Simulation) & Data/\allowbreak{}event & Effective event detection, Resource utilization \\
\cite{ref193} & FL~(C) & Collaborative secure data sharing (Generated) & Data & High accuracy; low latency \\
\cite{ref196} & FL~(C) & Parking space estimation (Birmingham) & Location & Accurate estimation \\
\cite{ref197} & FL~(C) & Power load prediction (Hangzhou charging) & Data & Under 3\% prediction loss \\
\cite{ref198} & FL~(C) & Energy prediction (Dundee charging) & Energy & Lowest RMSE 5.76\%; Reduced comm. cost \\
\cite{ref199} & FL~(C) & Station recommendation (Generated) & Data & +6\% AUC improvement; Improved generalization \\
\cite{ref200} & FL+\allowbreak{}BC~(D) & Multi-agent charging optimization (Dundee charging) & Vehicle/\allowbreak{}CS & Outperforms by 36--48\%; 89\% less comm. \\
\cite{ref201} & FL~(C) & Energy prediction (Generated) & Data & High probabilistic prediction accuracy \\
\cite{ref202} & FL~(S) & Decentralized route planning (mid-sized US city sim) & Network/\allowbreak{}route & Route optimization; Low latency \\
\cite{ref203} & FL~(S) & Route selection / traffic prediction (PeMS sim) & Data/\allowbreak{}task & Improved accuracy \\
\cite{ref204} & FL~(C) & Travel time estimation (Didi Chengdu/Xi'an) & Data/\allowbreak{}model & Improved accuracy/communication \\
\cite{ref205} & FL~(C) & Cross-area travel trajectory (Didi Chengdu/Xi'an) & Area/\allowbreak{}trajectory & Cross-area optimization \\
\cite{ref209} & FL~(C) & Client-server selection (MNIST/CIFAR-10/SVHN) & Data & Enhanced contextual accuracy \\
\cite{ref210} & FL~(C) & Resource management (Simulation) & Data/\allowbreak{}computation & Resource efficiency; reduced latency \\
\cite{ref211} & FL~(C) & Resource management (Simulation) & Agent & Efficient resource allocation \\
\cite{ref212} & FL~(S) & Model quality / 6G integration (BelgiumTSC) & Layer security & 96\% average accuracy \\
\cite{ref213} & FL~(C) & Anomalous event detection (Simulation) & Vehicle/\allowbreak{}Comm. & Enhanced reliability; 79\% data reduction \\
\cite{ref327} & FL~(D) & EV charging demand prediction (multiple EV datasets) & Data/\allowbreak{}aggregation & Lowest RMSE (0.4558); 67.87\% lower comm. \\
\cite{ref337} & FL~(D) & Reliable FL aggregator selection (Simulation) & Model & Secure aggregation; Up to 95\% accuracy \\
\cite{ref336} & FL~(C) & EV energy trading privacy (Generated) & Data & Reduced comm./comp. cost \\
\cite{ref342} & FL~(C) & EV charging occupancy prediction (Dundee EVCS) & Training/\allowbreak{}CS & Up to 97.2\% accuracy (10-min. ahead) \\
\cite{ref343} & FL~(C) & Battery SoC estimation (Musoshi/NASA/BMW) & Battery telemetry & Lowest RMSE/MAE, highest R\textsuperscript{2} \\
\cite{ref334} & FL~(D) & Cluster-based FL (MNIST / SUMO Berlin) & Data & About 85\% coverage; fastest convergence \\
\cite{ref339} & FL~(C) & Communication-efficient FL (AirComp sim) & Gradient & Up to +20\% accuracy; +50--70\% participation \\
\cite{ref340} & FL~(C) & Incentive mechanism (MNIST/GTSRB) & Data & About 85--88\% test accuracy; up to 512x less comm. \\
\cite{ref345} & FL~(C) & Communication-efficient training (MNIST/CIFAR-10) & Model & +33.23\% accuracy vs. FedCS \\
\cite{ref346} & FL~(S) & FL client selection / incentive (simulated auction) & Bidding data & +38.9\% platform utility; 31.3\% lower payments \\
\cite{ref344} & FL~(C) & Task offloading in VEC (Didi GAIA) & RSU & Higher completion rate/reward vs. baselines \\
\cite{ref348} & FL~(C) & Travel demand forecasting (Shanghai Didi) & Travel data & R\textsuperscript{2} up to 0.887 \\
\cite{ref349} & FL~(C) & Task offloading in VEC (Simulation) & Data & About 92\% task completion rate \\
\cite{ref353} & FL~(C) & Cross-camera vehicle re-ID (VeRi-776/VehicleID) & Data (image) & 56.1\%/60.0\% mAP; Reduced comp./bandwidth \\
\cmidrule{1-5}
\cite{ref214} & BC~(D) & IoV block-streaming service (Simulation) & User/\allowbreak{}edge-node & Improved hit rate; lower energy/delay \\
\cite{ref215} & BC~(D) & Toll pricing / traffic prediction (NY Thruway) & Data & 0.0012 MSE; 45.88ms, 53 bytes \\
\cite{ref216} & BC~(D) & Fair energy trading (Generated) & Transaction & 20ms verification; 6s latency \\
\cite{ref217} & BC~(D) & SDN controller security (Simulation) & Network control & Network/access control; High throughput; low latency \\
\cite{ref218} & BC~(D) & Secure network topology sharing (Generated) & Service & High throughput; minimized delay/energy \\
\cite{ref219} & BC+\allowbreak{}FL~(D) & Secure data sharing (Uber NYC/MNIST) & Model updates & Secure update;High model accuracy; fast convergence \\
\cite{ref220} & BC~(D) & Knowledge sharing (MNIST/CIFAR-10) & Layer-wise & +10\% better accuracy \\
\cite{ref325} & BC+\allowbreak{}FL~(D) & Cross-region vehicular data sharing (GTSRB-based) & Model & Limited data sharing; robust to 30--40\% poisoned nodes \\
\cite{ref341} & BC+\allowbreak{}FL~(D) & Trajectory-aware FL comm./incentive (LuST) & Model & On-chain recording; 0.92 acc.; fairness (Gini 0.177) \\
\cite{ref221} & BC~(D) & Secure SVM training (BCWD/ACAD) & Data/\allowbreak{}feature & Low time cost; High computation \\
\cite{ref222} & BC+\allowbreak{}FL~(D) & Secure vehicular communication (Generated) & Model/\allowbreak{}training & Optimal block arrival rate; minimized delay \\
\cite{ref195} & BC+\allowbreak{}FL~(D) & Message dissemination (Generated) & Message & High accuracy; +65.2\% faster, +8.2\% more efficient \\
\cite{ref318} & BC+\allowbreak{}FL~(D) & Trust management, large-scale IoV (simulated urban) & Trust records & Trusted decisions; 96.83\% trust prediction accuracy \\
\cite{ref319} & BC+\allowbreak{}FL+\allowbreak{}DP~(D) & Privacy-preserving data sharing (MNIST/CIFAR-10/SVHN) & Gradient & Comparable accuracy; 15.97x smaller uploads \\
\cite{ref320} & BC+\allowbreak{}FL~(D) & Lane-change prediction (NGSIM/MNIST/CIFAR) & Data & Low computation; High transaction throughput \\
\cite{ref321} & BC+\allowbreak{}FL+\allowbreak{}HE~(S) & Traffic analysis (MNIST/CIFAR-10/PeMS) & Model updates & Secure aggregation; Up to 96.4\%/95.5\% accuracy \\
\cite{ref322} & BC+\allowbreak{}FL+\allowbreak{}HE~(D) & Trust management / secure aggregation (CityPulse) & Dual-layered model & 100\% security effectiveness; 90ms latency \\
\cite{ref317} & BC+\allowbreak{}FL+\allowbreak{}HE~(S) & EV charging station recommendation (Dundee, Scotland) & Data & Within bounded error of central model; 5.2s faster \\
\cite{ref324} & BC~(D) & EV charging payment privacy (Ethereum testbed sim) & Identity, payments & Full anonymity/traceability; sub-ms sign/verify \\
\cite{ref326} & BC~(D) & UAV-assisted VEC resource allocation (Simulated) & Data & Near-optimal profit convergence \\
\cite{ref323} & BC+\allowbreak{}FL~(D) & Traffic flow prediction (DelDOT/METR-LA) & Model updates & TEE-protected decryption; 0.03s encryption \\
\cmidrule{1-5}
\cite{ref223} & HE~(D) & Matching for EV charging (Generated) & Location/\allowbreak{}identity & Optimal matching; Linear complexity \\
\cite{ref224} & HE~(S) & Dynamic pricing in V2G networks (Generated) & Price/\allowbreak{}k-threshold & Fair pricing; minimized overhead \\
\cite{ref225} & HE+\allowbreak{}FL+\allowbreak{}BC~(D) & EV charging demand prediction (Dundee/CAN) & Parameter & High prediction accuracy; low latency \\
\cite{ref226} & HE~(C) & Ride-sharing (Generated) & Route & High computation cost \\
\cite{ref227} & HE~(C) & Toll systems for congestion automation (UHF RFID) & Transponder & User control; Monetize driving data \\
\cite{ref338} & HE+\allowbreak{}FL~(C) & Multi-hop task offloading (Simulation) & Parameter & Highest throughput/revenue/satisfaction \\
\cite{ref354} & HE+\allowbreak{}DP~(D) & V2V EV energy trading matching (Simulated) & Coordinate & About 84\% match consistency (10 EVs) \\
\cite{ref355} & HE+\allowbreak{}BC~(D) & Data sharing in VANETs (Comma.ai/simulated) & Identity & Matches plaintext computation; 68.51\% lower delay \\
\cite{ref356} & HE~(S) & Verifiable data aggregation for V2G (simulated, 10k EVs) & Data & Near-100\% verification success; 608-bit payload \\
\cmidrule{1-5}
\cite{ref231} & SMPC+\allowbreak{}BC~(D) & Vehicular data protection (Generated) & Transaction & Higher security utility; Below 80\% accuracy \\
\cite{ref232} & SMPC~(N) & Secure vehicular path tracking (Beijing taxis) & Data/\allowbreak{}path & Prevents adversarial attack; numerical instability \\
\cite{ref233} & SMPC~(D) & CAV data validation (NYC Connected Vehicles Pilot) & Real-time/\allowbreak{}data & Secure storage; 7\textmu s delay (50 nodes) \\
\cite{ref234} & SMPC+\allowbreak{}HE~(D) & Collaborative filtering (Simulation) & User & Secure filtering; low time/comm. cost \\
\cite{ref358} & SMPC~(N) & EV charging station recommendation (Charlotte, NC) & Identity & Anonymous auth.; 53\% lower re-authentication cost \\
\cmidrule{1-5}
\cite{ref235} & DP~(C) & EV charging (ACN-Data) & User & Reduced interference/publishing errors \\
\cite{ref236} & DP~(C) & EV querying (OpenStreetMap) & Location & High QoS; ``privacy-for-free'' for majority \\
\cite{ref237} & DP~(C) & Parking / charging station (Santander) & Location/\allowbreak{}query & High accuracy; low overhead \\
\cite{ref238} & DP+\allowbreak{}FL~(S) & Traffic flow estimation (i-VANET) & Data & Resilient to inference/leakage attacks \\
\cite{ref194} & DP+\allowbreak{}FL~(C) & Robust VCPS (MNIST) & Model & Adversarial resistance; Low acc. drop at high budget \\
\cite{ref329} & DP~(D) & V2V energy trading matching (constructed EV dataset) & EV location & O(N) adaptive budget computation \\
\cite{ref332} & DP~(C) & EV charging scheduling (Simulated) & Arrival/\allowbreak{}departure & Stable high reward across budgets \\
\cite{ref333} & DP~(C) & Scalable EV charging control (IEEE test systems) & Arrival data & 2.01--3.06\% optimality gap \\
\cite{ref351} & DP+\allowbreak{}FL~(C) & EV battery SoH estimation (NASA/CALCE/SNL) & Data & Resist data leakage/attacks; Reduced time/comm. \\
\cite{ref352} & DP+\allowbreak{}FL~(C) & Energy consumption prediction (Vehicle Energy Data) & Data & Lower MAE than baselines \\
\cite{ref328} & DP~(C) & Personalized automotive data sharing (Synthetic) & Automotive data & Access control; RMSE less than uniform-budget LDP \\
\cite{ref331} & DP+\allowbreak{}FL~(C) & Privacy-preserving FL training (Adult/MNIST) & Data & Higher accuracy/faster convergence \\
\cite{ref330} & DP~(S) & Task offloading in VEC (Atari / Beijing sim) & Model/\allowbreak{}env. & Higher reward than DP-SGD baseline \\
\cite{ref350} & DP+\allowbreak{}FL~(S) & Task offloading in CVIS (Didi GAIA) & Task & 10.1\%/25.6\% lower delay; 13.1\%/4.4\% higher resource \\
\cite{ref347} & DP+\allowbreak{}FL~(C) & Driver expression recognition (FER2013/KMU-FED) & Gradient & Accuracy/F1 about 0.95; Low avg. delay \\
\cite{ref359} & DP~(D) & V2I comms./location (Santander parking) & Data & Faster key generation; Low response time \\
\end{longtable}
}

For collaborative task scheduling in vehicular edge computing, \cite{ref344} formulated offloading and resource allocation as a non-cooperative potential game and introduced Fed-MADDPG, in which federated aggregation of multi-agent actor-critic policies lets roadside units coordinate decisions without exchanging local state information. This improved task completion and reward distribution across varying traffic volumes and RSU computing capabilities. In the context of multi-server task offloading, \cite{ref349} combined FL with multi-agent deep RL so that vehicles, multi-access Edge Computing (MEC) servers, and the cloud jointly train offloading policies through a reward function balancing delay, energy, and privacy constraints without centralizing sensitive state information. For short-term travel demand forecasting on ride-hailing platforms, \cite{ref348} introduced a multi-factor federated ConvLSTM scheme that folds weather and points-of-interest data into the order-based feature space and applies a privacy-preserving sharing protocol to counter the accuracy loss induced by non-IID clients. The approach raised forecasting accuracy under both IID and non-IID conditions. A three-stage federated scheme for unsupervised domain adaptive vehicle re-identification distributed local re-ID model training across edge cameras and reconciled the resulting models on a central server through weight-regularized aggregation \cite{ref353}.

\subsubsection{Blockchain-based PPML}

There are hybrid frameworks combining BC and FL to enhance data security and resilience in VCPS. Study in \cite{ref214} proposed authentication scheme between vehicles and RSUs. Using on-chain hashing, off-chain integrity schemes, cryptographic algorithms, and certificate authentication, the system ensured anonymous service requests, two-way authentication, and privacy preservation. A hybrid approach in \cite{ref215} integrates BC for secure transactions between vehicle owners and regulatory bodies ensuring data reliability and payment transparency. To further optimize resources and transactions, a privacy-preserving energy trading scheme in \cite{ref216} used BC and ZKP, to ensure confidentiality in energy transaction between EVs and the power grid.

Authors in \cite{ref217} proposed a 5G-enabled fog computing paradigm where RSUs act as software-defined networking (SDN) controllers, managing BC operations and secure channel selection. This decentralized approach reduces reliance on central servers and implements reputation-scoring mechanisms for security. Similarly, \cite{ref218} presented dual-layered SDN-controlled vehicle edge computing (VEC) framework integrating BC for secure network topology sharing. By using enhanced PBFT algorithm, it improved throughput, reduced latency, and ensured data integrity.

To address trust and privacy of real-time data transmissions in IoV authors have proposed integrating BC with FL. In \cite{ref219}, a BC-empowered FL framework enabled distributed intelligence while preserving privacy. To further enhance the efficiency of knowledge sharing, authors in \cite{ref220} proposed a hierarchical BC framework combined with a layered FL approach. It consisted of multiple leader and player setup, where vehicles function as individual FL nodes. A lightweight Proof-of-Knowledge (PoK) consensus mechanism was utilized to mitigate BC overhead. To scale secure knowledge sharing across regions, \cite{ref325} proposed a hierarchical, two-layer BC architecture that partitions the IoV edge network into regions running asynchronous swarm learning in parallel. BC pruning combined with SVD-based model compression bounds the DAG ledger, and the resulting HB-MASL framework retains higher accuracy than several synchronous and asynchronous BC-FL baselines. Study \cite{ref341} mined real vehicular trajectories from the LuST simulation dataset to build a delay-weighted communication graph that drives client selection and score-weighted aggregation in FL, and layered this onto a BC-based smart-contract mechanism that records and rewards each node's multidimensional contribution. The resulting framework reports gains in convergence, communication stability, and reward fairness over baselines.

Privacy concerns in vehicular social networks are studied in \cite{ref221} through a secure SVM classifier training system based on BC and cryptographic techniques. It eliminated third-party intermediaries through smart contracts. A BC-powered autonomous FL system was proposed in \cite{ref222}, optimizing parameters like block size and arrival rate to enhance efficiency. Additionally, a BC-based FL solution for emergency message dissemination was introduced in \cite{ref195}. It leverages Proof-of-FL (PoFL) consensus to mitigate broadcast storms and low packet reception, and a Stackelberg game-based model to incentivize participation in model training.

Study \cite{ref318} embeds superposition-inspired neurons into a global vehicular trust model and combines this design with particle-swarm-optimized federated aggregation to counter the slow convergence in non-IID vehicular data. A BC ledger secured through hash functions and ZKPs anchors trust records, while a Shapley-value mechanism rewards vehicles for maintaining honest, high-quality reporting. Authors in \cite{ref319} targeted the communication overhead and single-point-of-failure risk by combining a federated DP scheme with a new Proof-of-Reputation-and-Learning consensus for the supporting BC. Study \cite{ref320} introduced a node selection mechanism that weighs vehicles' computing power, network conditions, and data utility before admitting them to a federated round, paired with a lightweight proof-of-knowledge consensus that keeps BC-backed aggregation practical for resource-limited roadside units. Study \cite{ref321} layered HE over a consortium-BC-secured federated pipeline so that roadside units can aggregate encrypted vehicle model updates without ever exposing them in plaintext, while a dynamic scaling factor adaptively allocates computing resources across connected vehicles. Authors in \cite{ref322} paired federated trust learning with a dual-layer RSA encryption so that model parameters stay confidential during transit and aggregation. A hybrid supervised and unsupervised anomaly detector flagged emerging attack patterns in vehicular data streams.

Authors in \cite{ref317} proposed a cloudlet-assisted recommender for EV CSs that trains a VFL model jointly over EV-side and station-side attributes while keeping the exchanged gradients and loss terms homomorphically encrypted. It replaced a single centralized aggregator and reduced the risk of tampering during recommendation. Study \cite{ref324} removed the trusted third-party billing entity from ICV charging by settling payments directly through a smart contract that validates ZKPs and an efficiency-improved ElGamal group signature scheme. Authors in \cite{ref326} formulated the resource trading among edge servers, unmanned aerial vehicles (UAVs), and vehicles in UAV-assisted vehicular edge computing as a three-tier Stackelberg game, coupling a proof-of-stake BC ledger with a multi-armed bandit reputation mechanism to steer edge servers toward trustworthy data sellers. The scheme jointly optimizes resource pricing and allocation under delay, energy, and security-level constraints. Study \cite{ref323} addressed the unbounded storage growth of BC-recorded federated updates by introducing a chameleon-hash-based redactable DAG BC whose rewriting rights are secured inside Intel SGX trusted execution environments rather than through costly secret-sharing cryptography. The resulting framework keeps encryption and aggregation overhead close to baseline while still permitting authorized on-chain data redaction.

\subsubsection{Homomorphic Encryption}

Existing literature has studied the use of HE to address privacy concerns in various IoV applications including charging location privacy, vehicle tracking prevention, and secure interactions within ride-sharing platforms. Study \cite{ref223} proposed a privacy-preserving distributed matching algorithm for EV charging. It leveraged the Paillier cryptosystem to secure location data. Similarly, \cite{ref224} introduced PADP, a privacy-preserving data aggregation and dynamic pricing scheme for V2G networks. It used HE to protect power consumption data while enabling real-time aggregation for dynamic pricing across regions. Additionally, \cite{ref225} developed a BC-based FL framework for EV charging demand prediction, using CKKS HE to ensure privacy during model training.

Several studies in the IoV ecosystem highlight the utility of HE for various services. For ride-sharing, \cite{ref226} proposed a framework using PHE to protect sensitive data while ensuring privacy-aware ridesharing and routing. To prevent vehicle tracking, \cite{ref227} introduced a privacy risk assessment model that evaluated risks associated with toll transponders. Study in \cite{ref228}, combined HE with BC for intelligent transportation, employing partially hashing HE and decryption (PHHE/D) for local data encryption at fog nodes. The secure, cost-optimal workload assignment (SCWA) algorithm ensured efficient processing, while the BC enhanced security and operational efficiency in the IoV network.

To enhance data security and confidentiality in V2X communication, several studies have integrated HE in the network. In \cite{ref229}, a privacy-aware intelligent forwarding solution is introduced for named data networking (NDN)-VANETs. It integrated HE with an enhanced Bayesian receiver forwarding decision (BRFD) mechanism to enable secure and private vehicle-to-base (V2B) data exchange. In \cite{ref104}, a decentralized privacy-preserving DL (DPDL) model for VANETs integrated FHE and BC to enable secure data exchanges among vehicles and edge nodes. It ensured data privacy and mitigated threats like model extraction.

Authors in \cite{ref338} jointly formulated multi-task FL and a cloud-edge Stackelberg incentive game to allocate scarce communication and computing resources. To guard uploaded parameters, the framework layers HE onto the federated aggregation step so the server can combine client updates without ever decrypting individual contributions. For location privacy in vehicle-to-vehicle energy trading among EVs, a two-layer matching scheme without ever observing plaintext locations was proposed in \cite{ref354}. The design explicitly quantifies the resulting privacy-utility tension through partial HE rather than a FHE scheme. A privacy computing scheme for vehicular ad hoc networks that pairs Paillier HE with a directed acyclic graph BC was proposed in \cite{ref355}. Addressing the trade-off between resource-constrained EVs and untrustworthy aggregators in vehicle-to-grid networks, study \cite{ref356} proposed an efficient aggregation scheme that adapts a multi-key HE message authentication code. The blind-signature pseudonyms and a split-authority masking design let the scheme resist forgery, replay, and collusion while keeping computation and communication overhead markedly low.

\subsubsection{Secure Multi-party Computing}

Studies utilizing SMPC focused on enhancing the efficiency and security of services like secure data sharing, communication, and personalized recommendation systems. For instance, the study \cite{ref231} introduced an AI-powered BC framework that combined SMPC with advanced cryptographic techniques to protect the privacy of vehicular data in ML applications. Another study in \cite{ref232} proposed a game-theoretic SMPC framework for privacy-preserving data sharing in the IoV. The framework also decentralized data collection through distributed servers and utilized spatio-temporal maps to ensure privacy while maintaining utility.

To validate mobility data on the IoV ecosystem, the authors in \cite{ref233} introduced the BELIEVE framework. It is also a BC-enabled framework that utilizes SMPC to ensure privacy-preserving real-time validation. It employed a privacy-by-design approach utilizing encrypted distance-based computations for mobility data validation through a Proof-of-Presence (PoP) consensus mechanism. Recommendation systems in IVI provide personalized content based on user preferences and behaviors. The study \cite{ref234} proposed a privacy-preserving multi-party collaborative filtering system for IVI recommendations, using SMPC and the Paillier cryptosystem. It employed a symmetric balanced block design for efficient aggregation which ensured sensitive data remains encrypted.

Authors in \cite{ref335} designed a tamper-resistant, physically unclonable function based mutual authentication scheme tailored to EV charging in V2G settings. Formal analysis confirmed resistance against impersonation, desynchronization, and machine-learning-based PUF modeling attacks. A dual-cloud framework for the IoVs was proposed in \cite{ref358}. It couples lightweight anonymous authentication with SMPC to deliver personalized CS recommendations without exposing user preferences or identity. The magazine article \cite{ref357} surveyed the security and privacy threats confronting EV in intelligent speed advisory systems.

\subsubsection{Differential Privacy}

For CAVs, smart charging and parking systems handle significant sensitive data, making them vulnerable to attacks and privacy breaches. Authors in \cite{ref235} proposed DP mechanism for protecting sensitive EV charging data in V2G networks, using sampling intervals and sliding windows. Another study \cite{ref236} proposed a privacy-preserving mechanism for EVs querying CSs. The authors propose approximate geo-indistinguishability, which provides two-fold privacy protection by safeguarding individual query locations and protecting against trajectory tracing in an online setting.

Many CSs are integrated with parking facilities, and while smart parking systems rely on continuous data sharing, they inherently pose risks to user privacy. Study \cite{ref237} proposed a privacy-preserving charging infrastructure system using ECC for mutual authentication and Laplace-distributed noise for LDP. Extending this to broader vehicle-infrastructure ecosystems, study \cite{ref238} addressed communication security between roadside infrastructure and vehicles by integrating FL with LDP in VANETs. Here, vehicle data is perturbed locally before being shared with external infrastructures, enabling collaborative model training while mitigating gradient leakage and inference attack. To further defend against adversarial attacks exploiting vehicle speed and location, authors in \cite{ref194} proposed a privacy-preserving VCPS. The system integrates FL with DP and applies layer-wise relevance propagation (LRP) to regulate perturbation values.

Authors in \cite{ref329} tackled the rigidity of static privacy budgets in V2V energy trading by proposing an adaptive privacy-preserving stable matching scheme that scales the DP budget to real-time EV density before applying Laplace or geo-indistinguishability perturbation to location data. Study \cite{ref332} designed a three-level multi-agent deep RL architecture in which discrete DP obfuscates the reserved arrival and departure times of EVs at a solar-powered CS. The layered design lets operators trade privacy strength against charging revenue and energy cost. Study \cite{ref333} tackled the feasibility problems that differential-privacy noise introduces into coordinated EV charging control by pairing a constrained RL policy with convex-hull-based action-space reduction, guaranteeing feasible dispatch under any admissible privacy-preserving realization of aggregate demand. To generalize state-of-health estimation across heterogeneous battery chemistries while resisting data leakage and adversarial noise, \cite{ref351} developed a federated framework that couples performance-induced client scheduling and discrepancy-weighted aggregation with Gaussian-mechanism perturbation of local updates. The method maintained low prediction error and outperformed conventional baselines. To curb the client drift arising from heterogeneous driving behavior in cross-device battery EV energy modeling, \cite{ref352} proposed RI-FedAvg, which regularizes local updates using a per-vehicle roughness index of the loss landscape, alongside a differentially private counterpart that layers Renyi-calibrated Gaussian noise onto the same regularization scheme. The roughness-aware variants lowered prediction error and converged in fewer rounds than the baselines.

Authors in \cite{ref328} addressed the rigidity of uniform privacy budgets in automotive data sharing by linking quantified access risk to individualized local DP budgets within a unified perturbation framework. Evaluation shows the risk-adaptive scheme markedly outperforms fixed global-budget local DP baselines while preserving user privacy preferences. Study \cite{ref331} embedded Gaussian differential-privacy noise into a federated stochastic primal-dual optimization procedure for IoV learning tasks. Another study \cite{ref330} coupled an online differentially private variant of double deep Q-networks with an autoencoder that obfuscates the vehicular state representation, targeting the leakage risk during offloading decisions in vehicular edge computing. Given the fluctuating privacy demands introduced by high vehicular mobility in cooperative infrastructure systems, \cite{ref350} proposed PA-FDRL, which pairs a federated double deep Q-network offloading policy with a dynamic DP budget. The adaptive noise-injection strategy kept leakage risk below a prescribed threshold while still reducing offloading delay and improving resource utilization.

Addressing the divergent privacy needs of public and personal vehicles, \cite{ref347} proposed a privacy-hierarchical offloading scheme for driver expression recognition. The non-private vehicles offload raw data under a soft actor-critic delay-aware policy while private vehicles train locally with differentially private stochastic gradient descent (SGD) before contributing only perturbed parameters to a dual-layer aggregation process. This improved global model accuracy without weakening the privacy guarantees. To remove third-party dependence from vehicle-to-infrastructure location sharing, \cite{ref359} anchored identification in a physical unclonable function combined with elliptic curve cryptography. Storing subsequent RSU records on an IOTA-based directed acyclic graph ledger rather than a conventional BC is presented as a route to quantum resistance and better scalability.

\subsection{Recent Advances}\label{sec:recent-advances}

We additionally review and discuss the recent advances spanning the three application domains and five PPML technique categories discussed above. Table~\ref{tbl:recent-advances} summarizes these studies, organized first by application domain and then by PPML technique, alongside a brief statement of each study's principal contribution. The following synthesizes the specific technical directions visible within each domain.

Within intelligent transportation and traffic management, the clearest shift is away from treating FL purely as a privacy wrapper around an otherwise standard forecasting model, and toward FL variants that are explicitly engineered for statistical and behavioral heterogeneity. Few studies are Personalized clustering to absorb cross-region spatio-temporal skew \cite{ref281}, fuzzy-inference client trust scoring to down-weight unreliable participants \cite{ref282}, and scene-specific multi-client training combined with attention-based aggregation \cite{ref283}. A parallel, and closely related, trend runs through the BC-based studies, which increasingly layer verifiable cryptographic primitives such as ZKPs for telemetry validity \cite{ref270}, attribute-based chameleon hashing for authorized ledger redaction \cite{ref269}, and constant-length ciphertext attribute-based encryption for crowdsourcing access control \cite{ref273}. Additionally, there are consensus mechanisms tailored to application latency, such as a dynamic-grouping PBFT variant for trajectory prediction \cite{ref277} and a fully decentralized two-tier microblock/keyblock scheme for sign recognition \cite{ref278}. This indicates BC is being used less as a passive audit log and more as an active substrate for encrypted, verifiable computation. SMPC and DP work in this domain has similarly moved from simple aggregation toward genuine multi-party coordination problems. Additive secret sharing recast as a matrix-completion task for arrival-profile estimation \cite{ref284}, two-server SMPC protecting a full recurrent inference pipeline for lane-change prediction \cite{ref285}, and gossip-based secret sharing coordinating platooning without a trusted broker \cite{ref286}. Together with the multi-key HE and secret-sharing scheme for poisoned-update filtering, \cite{ref280} explicitly target robustness to adversarial or corrupted participants alongside privacy, suggesting that privacy and Byzantine-resilience are increasingly being addressed jointly rather than as separate concerns in ITS research.

Within autonomous driving and safety-critical applications, adversarial robustness is the dominant and most consistent theme across every PPML technique family, reflecting the direct physical safety consequences of a compromised model in this domain. FL studies explicitly target Byzantine and gradient-based attacks through Gini-impurity detection \cite{ref298} and ensemble gradient-attack scoring combined with GNN-based reputation \cite{ref299}. BC studies pair reputation or trust scoring with consensus specifically for misbehavior and intrusion detection \cite{ref289,ref292,ref314}. The HE studies defend against poisoning through cosine-similarity update filtering \cite{ref303} and credibility-weighted incentive mechanisms \cite{ref312}. The SMPC and DP studies target false-data-injection defense \cite{ref316} and intrusion detection directly \cite{ref293,ref295,ref297}. A second, more pragmatic trend is visible in the HE studies, which favor selective or approximate encryption (such as CKKS applied only to the last layer \cite{ref303}, noise-resilient CKKS with dynamic precision scaling \cite{ref309}, and ring-topology modular decomposition \cite{ref310}) over full FHE, reflecting the real-time latency budget that autonomous driving imposes on any privacy mechanism. Perception-specific FL work has also emerged as a distinct sub-specialty, with techniques purpose-built for vision workloads such as camera-attentive positional embeddings for bird's-eye-view perception \cite{ref302}, lightweight adapter-based fine-tuning for 3D object detection \cite{ref304}, and selective client-update filtering \cite{ref308}, rather than reusing generic tabular or sequence-model FL pipelines. Finally, the SMPC study on privacy-preserving DL inference \cite{ref315} signals growing attention to protecting the full model lifecycle, extending beyond training-time privacy to the deployment and inference stage where safety-critical decisions are actually made.

Within communication infrastructure and smart services, electric-vehicle charging, battery management, and energy trading anchor the large majority of recent studies across every PPML technique (FL \cite{ref327,ref336,ref343}, BC \cite{ref317,ref324}, HE \cite{ref356}, SMPC \cite{ref358}, and DP \cite{ref328,ref333,ref351}). This confirms that the recent growth of PPML research in this domain is driven predominantly by EV infrastructure privacy needs rather than by smart services more broadly. A distinguishing characteristic of the BC studies here, relative to those in ITS and autonomous driving, is that BC is rarely used as a stand-alone privacy mechanism; instead it is consistently hybridized with another PPML technique that does the primary privacy work, such as VFL combined with additively homomorphic gradient aggregation \cite{ref317}, dual-layer RSA encryption paired with FL-based trust management \cite{ref322}, and trusted execution environment (TEE)-based secure aggregation layered onto a redactable DAG ledger \cite{ref323}, with BC itself contributing verifiability and accountability rather than confidentiality. Economic and game-theoretic mechanisms also appear far more prominently in this domain than in the other two, including Stackelberg pricing and multi-armed-bandit reputation for UAV-assisted resource allocation \cite{ref326}, a Stackelberg incentive game paired with HE aggregation for multi-hop task offloading \cite{ref338}, and online auction-based client selection \cite{ref346}, reflecting the inherently multi-stakeholder, incentive-driven nature of shared charging and energy infrastructure. Finally, the DP studies in this domain increasingly calibrate privacy budgets adaptively rather than applying a uniform global $\varepsilon$, whether through risk-aware personalized local DP budgeting for automotive data sharing \cite{ref328} or performance-induced client scheduling for battery state-of-health estimation \cite{ref351}. This is consistent with the personalization trend also visible in the ITS FL literature and points to heterogeneity-aware privacy calibration as a future direction.

\section{Challenges and Future Directions}\label{sec:challenges}

Each PPML technique trades privacy, computational overhead, and model utility differently, as summarized in Table~\ref{tbl:ppml-comparison}. The recent studies reviewed in this survey advanced non-IID robustness in FL, verifiable computation in BC, selective encryption in HE, dropout tolerance in SMPC, and adaptive budgeting in DP, while exposing a sharper and more specific open problem than the one they solved. Section~\ref{sec:challenges-techniques} discusses this progression in PPML techniques alongside the IoV-specific literature. Section~\ref{sec:challenges-domains} then delves into the application domains to discuss what makes privacy preservation hard, and in what direction that problem needs to be pushed.

{\scriptsize
\setlength{\tabcolsep}{3pt}
\begin{longtable}{@{}>{\raggedright\arraybackslash}p{0.055\linewidth}>{\raggedright\arraybackslash}p{0.055\linewidth}>{\raggedright\arraybackslash}p{0.045\linewidth}>{\raggedright\arraybackslash}p{0.25\linewidth}>{\raggedright\arraybackslash}p{0.115\linewidth}>{\raggedright\arraybackslash}p{0.41\linewidth}@{}}
\caption{Recent studies grouped by application domain and PPML technique.}\label{tbl:recent-advances}\\
\toprule
\textbf{Tech.} & \textbf{Ref.} & \textbf{Year} & \textbf{Application} & \textbf{Privacy target} & \textbf{Principal Contribution} \\
\midrule
\endfirsthead
\caption[]{Recent studies grouped by application domain and PPML technique. (continued)}\\
\toprule
\textbf{Tech.} & \textbf{Ref.} & \textbf{Year} & \textbf{Application} & \textbf{Privacy target} & \textbf{Principal Contribution} \\
\midrule
\endhead
\midrule
\multicolumn{6}{r}{\textit{Continued on next page}} \\
\endfoot
\bottomrule
\endlastfoot
\multicolumn{6}{@{}l}{\textbf{Domain I -- Intelligent Transportation and Traffic Management}} \\
\midrule
\multirow{4}{*}{FL} & \cite{ref279} & 2026 & Traffic flow/speed prediction & Data & MoE routing of recurrent/non-recurrent experts; syncs only utilized experts \\
 & \cite{ref281} & 2025 & Heterogeneity-aware traffic pred. & Data & Personalized clustering + time-window training for cross-region skew \\
 & \cite{ref282} & 2026 & Trust-aware forecasting & Data & Fuzzy-inference client trust scoring with temporal smoothing \\
 & \cite{ref283} & 2024 & Pedestrian trajectory pred. & Data & Federated multi-scene training with FedAtt aggregation \\
\cmidrule{1-6}
\multirow{7}{*}{BC} & \cite{ref269} & 2025 & Accident forensics & Accident records & Attribute-based chameleon hash for authorized ledger redaction \\
 & \cite{ref270} & 2026 & C-V2X data sharing & Data & zk-SNARK proof aggregation + LSTM reputation scoring \\
 & \cite{ref271} & 2025 & Emission-aware speed advisory & Data & Paillier-encrypted aggregation for mixed ICEV/EV fleets \\
 & \cite{ref272} & 2025 & Task offloading & Task data & FHE task preprocessing + potential-game decision algorithm \\
 & \cite{ref273} & 2026 & Crowdsourcing task matching & Worker/\allowbreak{}requester & Constant-length ciphertext ABE + searchable encryption \\
 & \cite{ref277} & 2025 & Trajectory prediction & Trajectory & Async. FL + BC dynamic-grouping PBFT \\
 & \cite{ref278} & 2026 & Sign recognition & Model & Fully decentralized two-tier microblock/keyblock consensus \\
\cmidrule{1-6}
\multirow{1}{*}{HE} & \cite{ref280} & 2025 & Secure aggregation & Model & Multi-key HE + secret sharing + poisoned-update filtering \\
\cmidrule{1-6}
\multirow{3}{*}{SMPC} & \cite{ref284} & 2025 & Arrival-profile estimation & Trajectory & Additive secret sharing + matrix completion \\
 & \cite{ref285} & 2025 & Lane-change prediction & Data & Two-server SMPC with secure tanh activation \\
 & \cite{ref286} & 2025 & Platooning coordination & Itinerary & Distributed secret sharing + gossip averaging \\
\cmidrule{1-6}
\multirow{3}{*}{DP} & \cite{ref274} & 2025 & Trajectory prediction & Parameter & Gaussian-noise FL parameter perturbation \\
 & \cite{ref275} & 2025 & Location publishing & Grid & Directional-threshold rotating group-leader selection \\
 & \cite{ref276} & 2024 & Trajectory publishing & Trajectory & LSTM-GAN synthesis + personalized density-aware DP \\
\midrule
\multicolumn{6}{@{}l}{\textbf{Domain II -- Autonomous Driving and Safety-critical Applications}} \\
\midrule
\multirow{6}{*}{FL} & \cite{ref298} & 2024 & CAV steering prediction & Data & Personalized local-round allocation + Gini-impurity Byzantine detection \\
 & \cite{ref299} & 2025 & AIoT road-vehicle security & Data & FGSA gradient-attack detection + GNN reputation scoring + RSMA scheduling \\
 & \cite{ref302} & 2023 & Bird's-eye-view perception & Position/\allowbreak{}model & Camera-attentive positional embeddings + adaptive multi-camera masking \\
 & \cite{ref304} & 2025 & 3D object detection & Data/\allowbreak{}parameter & Lightweight adapter-based local fine-tuning \\
 & \cite{ref305} & 2025 & AV group formation & Trajectory/\allowbreak{}sensing & Federated trust-score learning against collusive attacks \\
 & \cite{ref308} & 2025 & Real-time object detection & Sensor & YOLOv8 on-vehicle training + selective client-update filtering \\
\cmidrule{1-6}
\multirow{4}{*}{BC} & \cite{ref288} & 2025 & Privacy-preserving data sharing & Data & Dual DP + dynamic gradient aggregation + BC consensus \\
 & \cite{ref289} & 2024 & Driving distraction recognition & Data & Bayesian reputation-weighted aggregation + multi-sidechain PoQ \\
 & \cite{ref292} & 2025 & Collaborative intrusion detection & Data & TOP-K selection + SMOTE-ENN + Transformer BC FL \\
 & \cite{ref314} & 2025 & Misbehavior / intrusion detection & Model & FedAvg across MLP / LSTM / GRU + smart-contract model anchoring \\
\cmidrule{1-6}
\multirow{4}{*}{HE} & \cite{ref303} & 2026 & Robust FL against poisoning & Model & Cosine-similarity filtering + selective last-layer CKKS \\
 & \cite{ref309} & 2026 & Misbehavior / anomaly detection & Model updates & Noise-resilient CKKS + trust-weighted aggregation + PoA2 / IPFS \\
 & \cite{ref310} & 2026 & Cross-trust-domain learning & Model & GAT clustering/ring formation + modular-decomposition encryption \\
 & \cite{ref312} & 2025 & Collusion-resistant verifiable FL & Gradients & Homomorphic pseudorandom masking + credibility-weighted incentive \\
\cmidrule{1-6}
\multirow{3}{*}{SMPC} & \cite{ref313} & 2025 & Cross-manufacturer training & Proprietary data & Tripartite private comparison + constant-round MSB extraction \\
 & \cite{ref315} & 2025 & Privacy-preserving DL inference & Data/\allowbreak{}model & Tripartite comparison/multiplication + INT32 quantization \\
 & \cite{ref316} & 2025 & Traffic signal control & Occupancy data & Shamir secret-sharing + SecComp comparison + FDI defense \\
\cmidrule{1-6}
\multirow{3}{*}{DP} & \cite{ref293} & 2026 & Intrusion detection & Client/\allowbreak{}RSU & CNN-Transformer + dynamic time-varying DP noise \\
 & \cite{ref295} & 2026 & CAN bus intrusion detection & Data & Self-adaptive DP budget + FIM-guided gradient selection \\
 & \cite{ref297} & 2024 & In-vehicle/external intrusion detection & Model & Homogeneity-based slicing + Bi-LSTM + Laplace DP \\
\midrule
\multicolumn{6}{@{}l}{\textbf{Domain III -- Communication Infrastructure and Smart Services}} \\
\midrule
\multirow{6}{*}{FL} & \cite{ref327} & 2025 & EV charging demand/range pred. & Data/\allowbreak{}aggregation & Decentralized sync/async aggregation + adaptive comm. topology \\
 & \cite{ref334} & 2025 & Cluster-based FL & Data & Trust-aware cluster-head election + Kuhn-Munkres matching \\
 & \cite{ref336} & 2026 & EV energy trading privacy & Data & Federated logistic regression + IFTTA optimization \\
 & \cite{ref343} & 2025 & EV battery SoC estimation & Battery telemetry & Federated Adaptive Client Momentum aggregation \\
 & \cite{ref344} & 2025 & Task offloading in VEC & RSU & Federated MADDPG + potential-game reward shaping \\
 & \cite{ref346} & 2025 & Client selection/incentive & Bidding data & Online auction-based selection + reward function design \\
\cmidrule{1-6}
\multirow{7}{*}{BC} & \cite{ref317} & 2022 & EV charging station recommendation & Data & Vertical FL + additively HE gradients + BC cloudlets \\
 & \cite{ref320} & 2025 & Data sharing/lane-change pred. & Data & Dynamic node selection (FedANS) + lightweight PoK consensus \\
 & \cite{ref322} & 2025 & Trust mgmt./secure aggregation & Dual-layered model & Dual-layer RSA encryption + FL-based trust management \\
 & \cite{ref323} & 2026 & Traffic flow prediction, async. FL & Model updates & Redactable DAG BC + TEE-based secure aggregation \\
 & \cite{ref324} & 2025 & EV charging payment privacy & Identity, payments & ElGamal group signatures + zk-SNARK payment verification \\
 & \cite{ref326} & 2025 & UAV-assisted VEC resource alloc. & Data & Stackelberg pricing + MAB reputation + PoS BC \\
 & \cite{ref341} & 2026 & Trajectory-aware FL comm./incentive & Model & QoS graph modeling + communication-aware weighted aggregation \\
\cmidrule{1-6}
\multirow{2}{*}{HE} & \cite{ref338} & 2025 & Multi-hop task offloading & Parameter & Multi-task FL + Stackelberg incentive game + HE aggregation \\
 & \cite{ref356} & 2026 & Verifiable data aggregation, V2G & Data & Multi-key homomorphic MAC + ratcheted dual masking \\
\cmidrule{1-6}
\multirow{1}{*}{SMPC} & \cite{ref358} & 2026 & EV charging station recommendation & Identity & Dual-cloud secure two-party computation \\
\cmidrule{1-6}
\multirow{3}{*}{DP} & \cite{ref328} & 2026 & Personalized automotive data sharing & Automotive data & Risk-aware access control + personalized LDP budgeting \\
 & \cite{ref333} & 2025 & Scalable EV charging control & Arrival data & Constrained RL + convex-hull action-space reduction \\
 & \cite{ref351} & 2025 & EV battery SoH estimation & Data & Performance-induced client scheduling + Gaussian noise \\
\end{longtable}
}

\subsection{PPML Techniques}\label{sec:challenges-techniques}

\subsection{PPML Techniques}\label{sec:challenges-techniques}

This section discusses the challenges and future directions of the PPML techniques in light of the throughput, adversarial, and heterogeneity challenges of realistic IoV deployment scenarios.

\subsubsection{Federated Learning}

Statistical heterogeneity must be resolved without a central arbiter capable of imposing a corrective view of the data. A 93-study review finds no non-IID benchmarking method generalizes across applications, and that more than half of surveyed studies still default to MNIST or CIFAR-10 regardless of their stated target, undermining the reported robustness claims \cite{ref360}. IoV research showed some progress here. Early traffic-prediction FL treated heterogeneity as a secondary concern for a generic aggregation rule to absorb \cite{ref71,ref74}, whereas recent studies work around it directly with personalized clustering with time-window training to absorb cross-region skew \cite{ref281}, mixture-of-experts routing that synchronizes only the experts actually used \cite{ref279}, and fuzzy-inference client trust scoring \cite{ref282}. This is exactly the application-tailored, real-data-driven design \cite{ref360} argues the wider field still lacks, but these mechanisms remain validated on curated regional datasets rather than the standardized benchmarks. Systems heterogeneity compounds the problem. A 6G-focused survey finds a common on-device task can require 2\,GB of memory that many resource-constrained fleet devices cannot supply, and shows that FL's threat surface itself shifts under the hierarchical and peer-to-peer architectures increasingly used to manage this heterogeneity \cite{ref361}.

A largely separate challenge is that privacy and robustness defenses interact rather than compose. Poisoning can increase the effectiveness of gradient-inversion reconstruction, and colluding Sybil clients can bias a global model toward one participant's data, improving a subsequent membership-inference attack \cite{ref362}. This directly implicates the Byzantine detectors already reviewed in Section~\ref{sec:ad} \cite{ref298,ref299,ref303}, none of which is evaluated under simultaneous DP noise or secure aggregation. A benchmarking study spanning twelve widely used FL frameworks found that none of the frameworks implement geometric-median aggregation despite its documented robustness to poisoning \cite{ref362}.

\textit{Future Direction:} Closing this gap is likely to require evaluating existing defenses jointly with the privacy mechanisms they would realistically be deployed alongside, rather than validating each in isolation, and complementing them with federated unlearning that lets a fleet operator revoke a poisoned or voluntarily withdrawn contribution without full retraining \cite{ref360,ref361}. It is a capability that connects naturally to the privacy-aware federated RL formulations already emerging in the vehicular literature \cite{ref264}.

\subsubsection{Blockchain-based PPML}

Table~\ref{tbl:ppml-comparison} already shows BC scoring strongly on auditability but only conditionally on confidentiality. Transaction-graph analysis can deanonymize pseudonymous participants without breaking any cryptography, and because the ledger is immutable, a deanonymization technique after deployment can retroactively unmask a participant's entire history \cite{ref15}. It is a risk that the pseudonymity and traceable-identity schemes already used in IoV \cite{ref170} do not evaluate against directly. Consensus adds a second, better-quantified trade-off. No protocol jointly satisfies efficiency, security, and decentralization \cite{ref30}, and one IoV integration study measures consensus success collapsing from 97\% to 75\% as the malicious-node share rises from 10\% to 50\% \cite{ref320}. Recent work responds by layering verifiable cryptography onto lightweight consensus. These include chameleon-hash ledger redaction \cite{ref269}, aggregated zero-knowledge telemetry proofs \cite{ref270} and a two-tier microblock/keyblock structure \cite{ref278}. This is exactly the direction \cite{ref30} identifies as a research hotspot, though none of these schemes is load-tested near the malicious-node range where \cite{ref320} shows robustness actually collapsing.

Another challenge is coordinating FL specifically. On-chain storage does not scale, off-chain trusted computation "under the chain is extremely challenging" \cite{ref30}, and a DAG-based trust framework built to avoid a single ledger was found to introduce high overhead precisely because it multiplies validation points \cite{ref320}. The redactable-DAG-plus-SGX design already reviewed \cite{ref323} is the closest realized answer but remains a single deployment, not the general off-chain framework \cite{ref30} calls for. BC security also rests on non-quantum-safe signatures, and given vehicles' decade-plus service life, migration cannot realistically be deferred to a later update cycle the way it might be for shorter-lived consumer devices.

\textit{Future Direction:} The most immediate step is testing the verifiable-cryptography-on-lightweight-consensus direction \cite{ref269,ref270,ref278} at city scale and near the malicious-node range where \cite{ref320} shows robustness collapsing, rather than only at the single operating point each scheme was designed around. Alongside this, maturing TEE-secured off-chain verification beyond the single instance in \cite{ref323} (potentially by drawing on the TEE primitives already surveyed for other purposes \cite{ref247,ref254}) would address the general off-chain framework \cite{ref30} calls for. Lattice-based post-quantum signature migration paths deserve attention now given how long vehicles built on today's cryptographic assumptions will remain in service.

\subsubsection{Homomorphic Encryption}

Bootstrapping cost remains HE's core bottleneck. Noise grows exponentially under multiplication, and even FHE's roughly tenfold improvement over earlier bootstrapping leaves per-gate cost many orders of magnitude above a plaintext operation \cite{ref64}. No IoV study reviewed here solves this. Each shrinks what gets encrypted instead such as last-layer-only encryption \cite{ref303}, dynamic-precision CKKS \cite{ref309}, ring-topology modular decomposition \cite{ref310}. This is a pragmatic trajectory but an unevaluated one. Partial encryption leaves earlier gradients exposed to the very inversion attacks HE is meant to prevent, and no study quantifies the resulting protection gap relative to full encryption.

Multi-key FHE, the natural fit for a federated aggregation server, remains bounded to a fixed participant count and roughly doubles noise growth relative to single-key evaluation \cite{ref64} (opposite of a fleet's dynamically changing membership). Current work sidesteps this with HE/secret-sharing hybrids \cite{ref280} or authentication-only MACs \cite{ref356} rather than true multi-key FHE, and only few studies utilized HE in perception and object-detection tasks (Section~\ref{sec:recent-advances}). This is the clearest evidence that its overhead still rules out real-time vision workloads. Independent re-analysis has also found that bit-wise security parameters claimed by many HE libraries provide lower security levels in practice \cite{ref64}. This is a caution for any qualitative privacy rating, including Table~\ref{tbl:ppml-comparison}'s, that does not fix an implementation and parameter regime.

\textit{Future Direction:} Two directions seem most likely to move the needle. The first is characterizing how CKKS's precision-versus-depth trade-off \cite{ref363} should differ across the classification, regression, and detection heads common in IoV models, and quantifying what partial encryption schemes such as \cite{ref303,ref309} actually give up relative to full encryption. This is so far adapted at runtime within a single benchmark \cite{ref309}. The second is developing multi-key or threshold HE that supports an unbounded participant set and pairing hardware acceleration specifically with the convolutional and recurrent architectures that dominate IoV perception and forecasting, rather than pursuing general-purpose HE speedups. This echoes back to the adaptive, accuracy-versus-complexity balance \cite{ref363} identifies as HE's own closing priority.

\subsubsection{Secure Multi-party Computing}

SMPC's synchronous, multi-party structure introduces a weakest-link latency effect that FL and HE do not share. A two-server lane-change protocol is bottlenecked by whichever server is slowest \cite{ref285}, while protocols redesigned specifically for IoV comparison and activation operations report speedups of up to several hundred times over generic MPC frameworks \cite{ref313,ref315}. This shows that protocol genericity, not cryptography itself, drives most of the overhead. Privacy also rests on rigid two- or three-server non-collusion assumptions \cite{ref108,ref285,ref313,ref358} that fail categorically, not gracefully, if violated. IoV's dynamic RSU topology strains this classical fixed-party assumption, addressed so far only by \cite{ref307}'s dropout-tolerant packed secret sharing. It was a purpose-built response rather than a general theory applicable to the gossip-coordinated platooning and cross-carrier settings reviewed elsewhere \cite{ref284,ref286}.

SMPC also protects computation, not output. It does not stop a repeatedly queried result from leaking information over time, which is why several studies pair it with DP \cite{ref109,ref111}. Yet no study formally composes SMPC's guarantee with DP's budget accounting for a system queried continuously over an operational lifetime rather than trained once \cite{ref286,ref316}.

\textit{Future Direction:} Elastic protocols that avoid the weakest-link bottleneck documented in \cite{ref285}, and dynamic-membership SMPC that generalizes \cite{ref307}'s dropout tolerance beyond the single RSU-aggregation setting it was designed for, are the most tractable future steps. The harder but more consequential problem is formal SMPC-DP composition for continuous querying which is relevant to the always-on coordination settings in \cite{ref286,ref316}. The combination of SMPC with TEE-rooted trust \cite{ref247,ref254} could help by loosening the rigid non-collusion assumption it currently depends on.

\subsubsection{Differential Privacy}

DP-SGD's clip-then-noise mechanism means larger IoV models suffer proportionally larger utility loss \cite{ref364}. This is consistent with significant accuracy drop (87.76\% to 77.55\%) reported for gradient-level noise injection in FL-based trajectory prediction \cite{ref274}. The choice of composition theorem matters just as much. f-DP permits roughly 20$\times$ more training epochs than R\'enyi DP at the same budget \cite{ref16}, yet none of the FL/DP IoV studies reviewed specifies which theorem it uses. Under a worst-case adversary, empirically measured leakage has been shown to approach the full theoretical bound rather than staying safely below it \cite{ref364}, undercutting any assumption that higher-utility budgets are safely conservative.

DP's cost also falls unevenly. Gradient clipping mechanically penalizes outlier and underrepresented samples, with disparate impact documented even under mild 30--70\% class imbalance and loose budgets \cite{ref364}. This raises qualitative concern about sparse-region ITS bias with an actual mechanism. The DP also does not defend against poisoning, mirroring the privacy-robustness interaction already identified for FL \cite{ref16,ref362}. A systematic review of 106 DP-in-ML studies confirms fairness-aware budgeting remains unresolved field-wide and names adaptive DP combined with FL in VANET/IoV specifically as a priority direction \cite{ref365}.

\textit{Future Direction:} Reporting the composition theorem and the resulting round budget explicitly, rather than leaving it implicit, would be a low-cost step that makes IoV studies genuinely comparable. Beyond that, noise calibrated to data density or economic sensitivity rather than a single global budget, and the adaptive DP-plus-FL direction that \cite{ref365} names directly for VANET/IoV, are where this survey sees the most promising path forward. This would extend the geospatial and personalized calibration already explored for location publishing, automotive data sharing, and battery health estimation \cite{ref275,ref328,ref351}.

\subsection{PPML in IoV Application Domains}\label{sec:challenges-domains}

This section examines each application domain to identify its main privacy challenges and the directions needed to address them. Table~\ref{tbl:combined-matrix} summarizes the application areas considered across the domains and their coverage by different PPML techniques. The discussion is based on the studies reviewed for each application and focuses on practical challenges rather than the PPML techniques discussed in Section~\ref{sec:challenges-techniques}.

\subsubsection{Intelligent Transportation and Traffic Management}

Several of this domain's applications share a common shape. They consume continuous, fine-grained signals from many vehicles and are only useful if the resulting model or output is accurate and timely. \emph{Traffic flow, speed, and state prediction} \cite{ref71,ref281,ref282}, \emph{crowd flow and congestion monitoring} \cite{ref77,ref113}, and \emph{signal control and intersection management} \cite{ref83,ref109} all sit on this trade-off, but each strains it differently. Prediction and monitoring tasks need enough spatial and temporal resolution to be useful, so aggressive noise or aggregation directly erodes utility or accuracy. This tension is already visible in the accuracy loss reported when DP noise is added to gradients in traffic-flow FL \cite{ref112}. Because traffic density varies enormously across a city, a privacy setting tuned for a dense downtown is typically too weak or too strong everywhere else. Signal control adds a harder constraint on top. Decisions happen in a sub-second control loop, and the SMPC-plus-DP scheme already proposed for connected-vehicle signal control needed a dedicated two-stage stochastic program just to keep DP-induced noise from degrading control performance \cite{ref109}. This shows how much harder this constraint is to satisfy than for prediction or monitoring alone. Multiple stakeholders (a municipality, a private mobility operator, a neighbouring jurisdiction) may also need to coordinate control decisions without any one of them acting as a fully trusted central authority.

A second cluster concerns individuals rather than the road network. \emph{Vehicle positioning and trajectory prediction} \cite{ref87,ref283,ref277}, \emph{travel mode and behavior identification} \cite{ref91}, and \emph{location privacy, data publishing, and crowd sensing} \cite{ref107,ref111,ref275} all depend on data that is quasi-identifying almost by definition. A trajectory can reveal where someone lives and works even after coordinates are anonymized, and inferring a person's likely next location or travel mode is inferring something about their intent or habits. Publishing such data for downstream planning or research use adds a further tension, since a dataset useful enough for traffic engineers to plan around is often precise enough to re-identify the people in it. The density-aware, personalized perturbation already explored for trajectory publishing \cite{ref276} and the rotating-group-leadership scheme for location publishing \cite{ref275} both respond to exactly this tension, but each is tailored to one specific data type rather than the class of problem generally.

A differently shaped set of applications is \emph{security, trust, forensics, and attack resilience} \cite{ref99,ref269} and \emph{task offloading, resource allocation, and coordination} \cite{ref270,ref286}. Both involve multiple parties who need to cooperate without fully trusting one another. Forensic investigation requires sufficient information to identify a malicious actor after an incident. The redactable, authority-gated ledger proposed for accident forensics is one approach to balancing attribution with individual privacy \cite{ref269}. However, this creates a tension between identifying attackers and protecting the privacy of other network users. Resource sharing among mobility operators, edge providers, and RSUs presents a similar challenge. For example, the zero-knowledge C-V2X data-sharing scheme and the gossip-coordinated platooning framework reviewed earlier \cite{ref270,ref286} require sufficient transparency to support fair resource allocation. At the same time, they must prevent parties from revealing proprietary information about their demand or available capacity to competitors.

\begin{table}[htbp]
\centering
\scriptsize\normalfont
\setlength{\tabcolsep}{1.5pt}
\renewcommand{\arraystretch}{0.8}
\caption{PPML techniques across applications in the three domains. Hybrid studies appear under every technique they combine.}
\label{tbl:combined-matrix}
\begin{tabular}{
@{}
>{\raggedright\arraybackslash}p{0.15\linewidth}
>{\raggedright\arraybackslash}p{0.25\linewidth}
>{\raggedright\arraybackslash}p{0.20\linewidth}
p{0.13\linewidth}
p{0.10\linewidth}
p{0.13\linewidth}
@{}
}
\toprule
\textbf{Application area} &
\textbf{FL} &
\textbf{BC} &
\textbf{HE} &
\textbf{SMPC} &
\textbf{DP} \\
\midrule
\multicolumn{6}{@{}l}{\textbf{Domain I: Intelligent Transportation and Traffic Management}} \\
\midrule
Traffic Management &
\cite{ref71,ref72,ref73,ref74,ref75,ref76,ref279,ref281,ref282,ref94,ref102,ref112,ref83,ref84,ref77,ref78,ref86} &
\cite{ref93,ref94,ref95,ref96} &
\cite{ref73,ref96,ref102} &
\cite{ref109} &
\cite{ref72,ref94,ref112,ref109,ref77,ref113} \\
Mobility \& Behavior &
\cite{ref87,ref90,ref283,ref277,ref274,ref91} &
\cite{ref277} &
-- &
\cite{ref284,ref285} &
\cite{ref274,ref276} \\
Privacy \& Sensing &
-- &
\cite{ref273,ref107} &
\cite{ref105,ref106,ref107} &
\cite{ref108,ref111} &
\cite{ref111,ref114,ref115,ref275} \\
Security \& Trust &
\cite{ref98,ref101,ref278} &
\cite{ref69,ref98,ref99,ref100,ref101,ref269,ref278,ref104} &
\cite{ref98,ref104,ref280} &
\cite{ref280} &
\cite{ref98,ref280} \\
Resource Management &
\cite{ref85} &
\cite{ref270,ref271,ref272} &
\cite{ref271,ref272,ref103} &
\cite{ref110,ref286} &
\cite{ref103} \\
\midrule
\multicolumn{6}{@{}l}{\textbf{Domain II: Autonomous Driving and Safety-critical Applications}} \\
\midrule
Driver/vehicle Control &
\cite{ref121,ref123,ref124,ref126,ref177,ref118,ref129,ref130,ref134,ref135,ref136} &
-- &
\cite{ref124,ref175,ref177} &
\cite{ref179,ref181,ref316} &
\cite{ref182} \\
Decision-Making &
\cite{ref139,ref141,ref142,ref143,ref144,ref145,ref146,ref298,ref301,ref305} &
\cite{ref166,ref167,ref168} &
\cite{ref142} &
-- &
\cite{ref183,ref184,ref185} \\
Perception &
\cite{ref153,ref154,ref155,ref156,ref157,ref158,ref159,ref160,ref161,ref162,ref164,ref165,ref302,ref304,ref306,ref308} &
-- &
-- &
-- &
-- \\
Security \& Trust &
\cite{ref299,ref169,ref172,ref173,ref174,ref287,ref288,ref289,ref292,ref314,ref300,ref303,ref309,ref310,ref311,ref312,ref180,ref188,ref293,ref294,ref295,ref296,ref297} &
\cite{ref169,ref170,ref171,ref172,ref173,ref174,ref287,ref288,ref289,ref290,ref291,ref292,ref314,ref309,ref189} &
\cite{ref170,ref290,ref300,ref303,ref309,ref310,ref311,ref312,ref307} &
\cite{ref180,ref307,ref313,ref315} &
\cite{ref171,ref288,ref187,ref188,ref189,ref293,ref294,ref295,ref296,ref297} \\
\midrule
\multicolumn{6}{@{}l}{\textbf{Domain III: Communication Infrastructure and Smart Services}} \\
\midrule
Energy \& Charging &
\cite{ref197,ref198,ref199,ref200,ref201,ref327,ref336,ref342,ref343,ref317,ref225,ref351,ref352} &
\cite{ref200,ref216,ref317,ref324,ref225} &
\cite{ref317,ref223,ref224,ref225,ref354,ref356} &
\cite{ref358} &
\cite{ref354,ref235,ref236,ref329,ref332,ref333,ref351,ref352} \\
Mobility Services &
\cite{ref196,ref202,ref203,ref204,ref205,ref348,ref323} &
\cite{ref215,ref323} &
\cite{ref227,ref226} &
-- &
\cite{ref237} \\
Resource Management &
\cite{ref209,ref210,ref211,ref212,ref334,ref337,ref339,ref340,ref344,ref345,ref346,ref349,ref338,ref350} &
\cite{ref326} &
\cite{ref338} &
-- &
\cite{ref330,ref350} \\
Security \& Trust &
\cite{ref190,ref191,ref192,ref193,ref213,ref219,ref318,ref319,ref320,ref321,ref322,ref325,ref341,ref238,ref194,ref331} &
\cite{ref214,ref217,ref218,ref219,ref220,ref221,ref222,ref195,ref318,ref319,ref320,ref321,ref322,ref325,ref341,ref355,ref231} &
\cite{ref321,ref322,ref355,ref234} &
\cite{ref231,ref232,ref233,ref234} &
\cite{ref319,ref238,ref194,ref328,ref331} \\
Identity/\allowbreak personalization &
\cite{ref353,ref347} &
-- &
-- &
-- &
\cite{ref347,ref359} \\
\bottomrule
\end{tabular}
\end{table}

\textit{Future Direction:} Progress in this domain requires privacy mechanisms that can adapt to local traffic conditions and time of day. This could extend the density-aware calibration demonstrated for location publishing \cite{ref275} to prediction and monitoring tasks. Another important direction is conditional or selective disclosure. Identity or proprietary information should only be revealed when specific and auditable conditions are met. This can build on the redaction and access-control approaches explored in forensics and crowdsourcing \cite{ref269,ref273}. City-scale evaluation is also needed for the verifiable and lightweight consensus schemes explored in this domain \cite{ref269,ref270,ref278}. These approaches have not yet been tested at the scale required for metropolitan deployments. Additionally, UAV-assisted and edge-collaborative sensing remain important research gaps. These architectures extend data collection beyond fixed roadside infrastructure and have not yet been sufficiently addressed in the ITS literature.

\subsubsection{Autonomous Driving and Safety-critical Applications}

This domain includes applications that differ in how directly they affect physical safety. \emph{Driver and occupant monitoring} \cite{ref121,ref124,ref177} includes drowsiness, distraction, fatigue, and activity recognition. These applications collect highly sensitive biometric and behavioral data continuously inside a private space. However, this information must also be processed quickly to detect dangerous conditions and allow timely action. The HE- and FL-based drowsiness and fatigue detection methods reviewed earlier showed that encryption can introduce additional computation and communication overhead compared with fully local models \cite{ref124,ref177}. Therefore, achieving strong privacy while maintaining very low latency is particularly challenging. \emph{Steering, motion control, and speed or signal control} \cite{ref118,ref134,ref179} face a similar challenge. Privacy mechanisms used in real time control systems must not introduce delays that could affect vehicle operation. For example, the affine-masked cooperative cruise-control scheme keeps its overhead below 30\,ms to meet this requirement \cite{ref179}. Additionally, these safety-related decisions need sufficient records for post-incident analysis, which can conflict with minimizing stored data.

\emph{Trajectory prediction and publishing} \cite{ref141,ref298} and \emph{vehicle positioning and decision-making} \cite{ref166} extend the privacy concerns of individual tracking into cooperative perception. Sharing a predicted trajectory with nearby vehicles can help prevent collisions, but it also reveals information about a vehicle's location and intended movement to potentially untrusted parties. In addition, safety-related decisions need to remain auditable even when they are performed using encryption or secret sharing. This creates a trade-off between protecting information and maintaining transparency for later verification. \emph{Perception}, including object, road, sign, and scene detection \cite{ref153,ref302,ref308}, presents a particularly important challenge in this domain. It is the largest and most active application area in the reviewed literature, but most existing approaches rely on FL. This is because processing high-volume video and LiDAR data with convolutional models is difficult to combine with cryptographic PPML under strict real-time requirements. As a result, current privacy protection for perception mainly relies on system architecture, such as keeping raw sensor data on the device, as demonstrated by the federated bird's-eye-view and adapter-based detection frameworks reviewed earlier \cite{ref302,ref304}.

\emph{Security: misbehavior, intrusion, attack, and trust} \cite{ref298,ref303,ref309} is a common requirement across nearly all applications in this domain. The Byzantine and poisoning defenses reviewed earlier, including Gini-impurity-based client filtering \cite{ref298} and cosine-similarity-based update screening under selective HE \cite{ref303}, are mainly evaluated against specific attack scenarios. They do not fully address attackers that can adapt after learning how the defense operates. Therefore, security remains an ongoing challenge rather than a problem with a single final solution.

\textit{Future Direction:} The main direction for this domain is to design privacy mechanisms that can operate together with continuous and adaptive security monitoring. This includes extending the trust-scoring and anomaly-detection approaches already demonstrated \cite{ref298,ref299,ref309} to address attackers that specifically target the detection mechanisms, rather than only the underlying model. Privacy mechanisms must also maintain the auditability required for vehicle safety. More specifically, future work should evaluate how much privacy selective-encryption approaches such as \cite{ref303,ref309} provide compared with full encryption. Cryptographic protocols designed specifically for convolutional perception models are also needed, rather than adapting general-purpose HE or SMPC methods. Finally, privacy-preserving explainability methods should be developed that are compatible with safety and security requirements. Existing tools such as saliency-map inspection \cite{ref263} may not be directly applicable when inference is performed using encryption or secret sharing.

\subsubsection{Communication Infrastructure and Smart Services}

The applications in this domain share an important feature that is less prominent in the previous two domains. They often involve financial transactions between organizations rather than individuals. \emph{EV charging, energy trading, and battery management} \cite{ref197,ref317,ref343} are clear examples. Usage data has commercial value because it can reveal demand patterns to competitors or utilities. The participants, such as charging operators, aggregators, and utilities, may also have competing interests. Therefore, privacy mechanisms must support cooperation while protecting commercially sensitive information. They must also support verifiable billing, since keeping a transaction private is not sufficient if its fairness cannot be verified. The VFL and homomorphic-aggregation scheme proposed for charging-station recommendation demonstrates this need \cite{ref317}. \emph{Parking, toll, and congestion pricing} \cite{ref196,ref237} face a similar billing challenge. Payments usually need to be linked to an accountable identity, making complete anonymity impractical. The main goal is therefore to verify payment or eligibility without revealing the user's complete movement history, as demonstrated by the ECC and local-DP parking-recommender scheme reviewed earlier \cite{ref237}.

\emph{Route planning, travel-time estimation, and ride-hailing} \cite{ref202,ref204,ref348} involve the same location and trajectory privacy concerns discussed for ITS. However, these applications often involve a commercial platform acting as a semi-trusted intermediary. Competing ride-hailing operators may also benefit from sharing data but have limited incentives to do so. \emph{Resource management, task offloading, and edge or network optimization} \cite{ref209,ref344,ref326} face a similar challenge. Fair allocation of computing resources and bandwidth requires information about the available capacity of different providers. However, providers may not want to reveal their operational or pricing information to competitors. This is also present in the UAV-assisted resource-trading scheme \cite{ref326}.

Two other applications are also important in this domain. \emph{Security, anomaly detection, trust, and data or knowledge sharing} \cite{ref318,ref322} involve the same cross-organizational trust challenge seen in ITS. These systems need to identify malicious actors while protecting legitimate participants, often across organizations without a common trusted authority. Current solutions mainly use combinations of blockchain and encryption that are designed for individual applications \cite{ref317,ref322,ref323}. \emph{Vehicle re-identification and driver-facing applications} \cite{ref353,ref359} present a different challenge. Their main purpose is to recognize returning vehicles or drivers and provide continuous services. However, re-identification is also a direct privacy concern. Therefore, the privacy challenge is part of the application itself rather than an external issue.

\textit{Future Direction:} Overall, the applications in this domain highlight several important research needs. Privacy mechanisms should consider not only personal privacy but also the economic value and commercial sensitivity of data. This could extend the risk- and performance-aware privacy budgeting approaches explored for automotive data sharing and battery health estimation \cite{ref328,ref351} to charging and energy-market data. Standardized and interoperable privacy schemes are also needed to allow organizations with different levels of trust to cooperate without developing separate solutions for each application. This would improve upon the application-specific blockchain and confidentiality approaches reviewed so far \cite{ref317,ref322,ref323}. For identity-dependent services, privacy-by-design approaches should focus on limiting the information exposed during re-identification rather than eliminating re-identification completely. This is more realistic because re-identification is necessary for these services. Similar considerations will also be important for UAV-assisted coordination \cite{ref326}, where mobile aerial nodes introduce additional mobility and trust challenges compared with traditional ground-based RSU systems.

\section{Conclusion}\label{sec:conclusion}

Integration of ML into the IoV has significantly enhanced transportation efficiency and autonomous driving capabilities. However, it also introduces privacy risks due to the sensitivity of vehicular, environmental and user data. Architectural PPML techniques such as FL and BC-PPML enable privacy preserving decentralized collaboration, while computational PPML techniques such as HE, SMPC and DP protect data confidentiality and limit unauthorized inference, though robustness to adversarial and poisoning attacks remains a largely open challenge. This survey provided a comprehensive review of the recent advances in adopting PPML techniques for IoV applications. We systematically analyzed the privacy challenges inherent to ML-driven IoV systems, including sensitive data exposure, adversarial attacks, and communication vulnerabilities. We categorized IoV applications into three key domains and evaluated how PPML techniques effectively mitigate privacy risks while preserving utility. Despite various advancements, significant challenges remain, such as balancing privacy-utility trade-offs, managing computational overhead, and ensuring scalability across heterogeneous networks. To overcome these, we further discussed potential future directions such as hybrid PPML framework and lightweight encryption for edge devices, among others.

\section*{Acknowledgments}
This work was partially supported by the Canada Research Chairs Program and the Natural Sciences and Engineering Research Council of Canada (NSERC).

\bibliographystyle{unsrtnat}
\bibliography{references}

\end{document}